\documentclass[11pt, draftcls, onecolumn]{IEEEtran}

\usepackage{amsfonts,amsmath,amssymb,graphicx,epstopdf,cite,subfigure,psfrag,epsfig,amscd}

\newtheorem{theorem}{Theorem}[section]
\newtheorem{lemma}[theorem]{Lemma}
\newtheorem{definition}[theorem]{Definition}

\begin{document}


\title{Greedy Sparse Signal Recovery with Tree Pruning}

\author{\IEEEauthorblockN{Jaeseok Lee, Suhyuk Kwon, Jun Won Choi, and Byonghyo Shim\\}

\thanks{J. Lee, S. Kwon and B. Shim are with Dept. of Electrical and Computer Engineering, Seoul National University, Seoul, Korea, and J. Choi is with Dept. of Electrical Engineering, Hanyang University, Seoul, Korea.

This work was sponsored by Communications Research Team (CRT), DMC R$\&$D Center, Samsung Electronics Co. Ltd, the MSIP (Ministry of Science, ICT $\&$ Future Planning), Korea in the ICT R$\&$D Program 2013 (KCA-12-911-01-110) and the NRF grant funded by the Korea government (MEST) (No. 2012R1A2A2A01047510).

A part of this paper was presented in International Symposium on Information Theory (ISIT) 2014.}

}

\maketitle

%
\begin{abstract}

%
Recently, greedy algorithm has received much attention as a cost-effective means to reconstruct the sparse signals from compressed measurements. Much of previous work has focused on the investigation of a single candidate to identify the support (index set of nonzero elements) of the sparse signals.
Well-known drawback of the greedy approach is that the chosen candidate is often not the optimal solution due to the myopic decision in each iteration.
In this paper, we propose a greedy sparse recovery algorithm investigating multiple promising candidates via the tree search.
Two key ingredients of the proposed algorithm, referred to as the matching pursuit with a tree pruning (TMP), to achieve efficiency in the tree search are the {\it pre-selection} to put a restriction on columns of the sensing matrix to be investigated and the {\it tree pruning} to eliminate unpromising paths from the search tree.
In our performance guarantee analysis and empirical simulations, we show that TMP is effective in recovering sparse signals in both noiseless and noisy scenarios.
\end{abstract}
%
%


\section{Introduction}

In recent years, compressive sensing (CS) has received much attention as a means to recover sparse signals in underdetermined system \cite{candes2006robust,candes2005decoding,liu2010orthogonal,candes2008restricted,tibshirani1996,dantzig,caiomp,wang2012recovery,needell2009cosamp,dai2009subspace,raginsky2010compressed,gao2011compressive,khajehnejad2011sparse}.
Key finding of the CS paradigm is that one can recover signals with far fewer measurements than traditional approaches use as long as the signals to be recovered are sparse and the sensing mechanism roughly preserves the energy of signals of interest.

It is now well known that the problem to recover the sparest signal $\mathbf{x}$ using the measurements $\mathbf{y} = \mathbf{\Phi x}$ is formulated as the $\ell_0$-minimization problem
%
\begin{eqnarray}
\label{eq:ell0min}
\min_\mathbf{x} \| \mathbf{x} \|_0 \,\,\,\,\,\,\,\, \text{subject to} \,\,\, \mathbf{y} = \mathbf{\Phi x}
\end{eqnarray}
where $\mathbf{\Phi} \in \mathbb{R}^{M \times N}$ is often called sensing matrix.
Since solving this problem is combinatoric in nature and known to be NP-hard\cite{candes2006robust}, early works focused on the $\ell_1$-relaxation method, such as Basis Pursuit (BP)\cite{candes2006robust}, BP denoising (BPDN) \cite{chen1999atomic} (also known as Lasso \cite{tibshirani1996}), and Dantzig selector \cite{dantzig}.
Another line of research receiving much attention in recent years is a greedy approach. In a nutshell, greedy algorithm attempts to find the support (index set of nonzero entries) in an iterative fashion, returning a sequence of estimates of the sparse input vector.
Although the greedy algorithm, such as orthogonal matching pursuit (OMP) \cite{caiomp}, is relatively simple to implement and also computationally efficient, performance is in general not so appealing, in particular for the noisy scenario.

\begin{figure}[t]
\begin{center}
		\includegraphics[width=135mm]{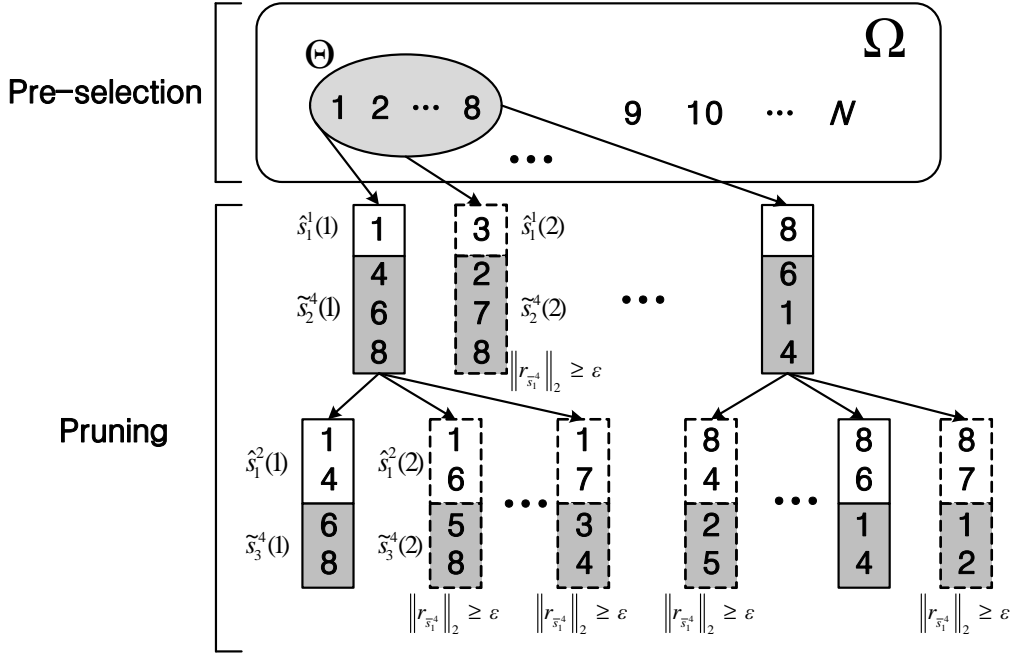}
		\caption{Illustration of the proposed TMP algorithm. Path with dotted box is pruned from the tree since the magnitude of the residual is larger than the threshold $\epsilon$.}\label{fig:big_pic}
\end{center}
\end{figure}

The aim of this paper is to introduce an efficient tree search algorithm to recover the sparse signal referred to as the \emph{matching pursuit with a tree pruning} (TMP). Our approach significantly reduces the computational burden of the exhaustive search yet achieves excellent recovery performance in both noiseless and noisy scenarios.
%
Two key ingredients of the TMP algorithm accomplishing this mission are the {\it pre-selection} to put a restriction on columns of $\mathbf{\Phi}$ to be investigated and the {\it tree pruning} to eliminate unpromising paths from the search tree.
In the pre-selection stage, we choose a small number of promising columns in the sensing matrix using a conventional greedy algorithm. If we denote the set of column indices obtained in the pre-selection stage as $\Theta$, then we set $N \gg | \Theta | > K$ where $K$ is the sparsity of the input vector ($\|\mathbf{x}\|_0 = K$).
When we construct the tree for the search, we only use elements of $\Theta$ as a child node in the branching process so that relentless growth of the tree can be prevented.
In our empirical results, we show that TMP achieves near best performance only with $| \Theta | \approx 2K$.
Once the pre-selection is finished, a tree search is performed to find best estimate of support using the pre-selected set $\Theta$.
As mentioned, when we select the child node (new estimate of the support element), we only consider the elements of $\Theta$.
As a result, the number of all possible paths in the tree is reduced from $\binom{N}{K}$ to $\binom{| \Theta |}{K}$. While this reduction is phenomenal, searching all of these is still computationally demanding, in particular for large $N$ and nontrivial $K$.
In order to alleviate computational burden and at the same time maintain the effectiveness of the search, we introduce an aggressive tree pruning strategy by which unpromising paths are removed from the tree.
Note that to perform the tree pruning, we need to compare the cost function $J( \Lambda ) = \| \mathbf{y} - \mathbf{\Phi}_\Lambda \hat{\mathbf{x}}_\Lambda \|_2$ of the full-blown candidate $\Lambda$ ($\|\Lambda\|_0=K$) against a pruning threshold.
However, direct evaluation of $J( \Lambda )$ is not possible in the middle of search due to the causality of the search process so that we combine already selected indices (henceforth dubbed as the \emph{causal set}) and roughly estimated indices (\emph{noncausal set}).
If this {\it roughly} estimated cost function is greater than the deliberately designed pruning threshold $\epsilon$ (i.e., $J( \Lambda ) > \epsilon$), further investigation of the path is hopeless and hence we prune the path from the tree immediately.
%

In our analysis, we show that the proposed method can accurately identify the support of $K$-sparse signal and hence reconstruct the original sparse signal accurately in the noiseless setting if the sensing matrix satisfies the property so called restricted isometry property (RIP) (Theorem \ref{thm_noiseless_suff_cond1}).
In the noisy setting, we show that the accurate identification of support is possible if the signal power is sufficiently larger than the noise power (Theorem \ref{thm_noisy_success}).
%
%
In our empirical simulations, we confirm that TMP performs close to an ideal estimator\footnote{The estimator that has a prior knowledge on the support (which component of the sparse vector is zero or not) is often called {\it Oracle estimator}.} (often called Oracle estimator \cite{chan2013projection}) in the high SNR regime.

The rest of this paper is organized as follows. In Section \ref{sec:tmp}, we introduce the proposed TMP algorithm.
In Section \ref{sec:analysis}, we analyze the recovery condition under which TMP identifies the support accurately in the noiseless and noisy scenarios.
%
%
In Section \ref{sec:simulation}, we provide the empirical results and then conclude the paper in Section \ref{sec:conclusion}.

%
\begin{table}[t]
\begin{center}
\centering
\caption{The TMP algorithm}
\label{tab:alg}
\begin{tabular}{lr}
\hline
\hline
\multicolumn{2}{l}{\textbf{Input:} measurement $\mathbf{y}$, sensing matrix $\mathbf{\Phi}$, sparsity $K$, initial threshold $\epsilon_1$}	 \\
\multicolumn{2}{l}{\textbf{Output:} Estimated signal $\hat{\mathbf{x}}$ } \\
\multicolumn{2}{l}{\textbf{Initialization:} $i:=0$, $S^0 := \emptyset$ } \\
\hline
$\Theta = f_\text{preselection} \left( \mathbf{y},\, \mathbf{\Phi},\, p \right)$ & ({\it preselection}) \\
\multicolumn{2}{l}{{\bf while} {$i < K$} {\bf do} } \\
\multicolumn{2}{l}{ \hspace{2mm}$i := i+1$, $S^{i} :=\emptyset$, $\epsilon_{i+1} := \epsilon_i $ } \\
\multicolumn{2}{l}{ \hspace{2mm}{\bf for} {$ l = 1$ {\bf to} $| S^{i-1} | $} {\bf do} } \\
\multicolumn{2}{l}{ \hspace{2mm} $\theta := \Theta \setminus \hat{s}_1^{i-1} (l)$} \\
\multicolumn{2}{l}{ \hspace{2mm} {\bf for} {$ j = 1 $ {\bf to} $|\theta|$} {\bf do} }\\
					\hspace{6mm} $\hat{s}_1^i := \hat{s}_1^{i-1} (l) \cup \left \{ s_i (j) \right \}$ & ({\it update $j$-th path}) \\
					\hspace{6mm} {\bf if} {$\hat{s}_1^i \not \in S^i$} {\bf then} & ({\it check the duplicated path}) \\
					\hspace{8mm} $\tilde{s}_{i+1}^K \!=\! \arg \! \mathop{\max} \limits_{ \mathop{s \subset \Omega ,} \limits_{|s| = K-i}} \! \| \mathbf{\Phi}_{s}' \!\mathbf{r}_{\hat{s}_1^i} \!\! \|_2$ & ({\it support estimation}) \\
\multicolumn{2}{l}{ \hspace{8mm} $\bar{s}_1^K = \hat{s}_1^i \cup \tilde{s}_{i+1}^K$, $\mathbf{r}_{\bar{s}_1^K} = \mathbf{P}_{\bar{s}_1^K}^\bot \mathbf{y}$} \\
					\hspace{8mm} {\bf if} $\| \mathbf{r}_{\bar{s}_1^K} \|_2 \leq \epsilon_i$ {\bf then} & ({\it pruning decision}) \\
\multicolumn{2}{l}{ \hspace{10mm} $S^{i} := S^{i} \cup \hat{s}_1^i$, $I^* := \bar{s}_1^K$} \\
\multicolumn{2}{l}{ \hspace{10mm} {\bf if} $\| \mathbf{r}_{I^*} \|_2 \leq \epsilon_{i+1}$ {\bf then} } \\
					\hspace{12mm} $\epsilon_{i+1} := \left\| \mathbf{r}_{I^*} \right\|_2$ & ({\it update pruning threshold}) \\
\multicolumn{2}{l}{ \hspace{10mm} {\bf end if} }\\
\multicolumn{2}{l}{ \hspace{8mm} {\bf end if} }\\
\multicolumn{2}{l}{ \hspace{6mm} {\bf end if} }\\
\multicolumn{2}{l}{ \hspace{4mm} {\bf end for} }\\
\multicolumn{2}{l}{ \hspace{2mm} {\bf end for} }\\
\multicolumn{2}{l}{{\bf end while} }\\
{\bf return} $\hat{\mathbf{x}}^* = \mathbf{\Phi}^{\dagger}_{I^*} \mathbf{y}$ & ({\it signal reconstruction}) \\
\hline
\end{tabular}
\end{center}
The $f_\text{preselection}(\cdot)$ is a function to choose multiple promising indices (see Section II.A).
\end{table}
%

\section{Matching pursuit with a tree pruning}
\label{sec:tmp}

The proposed TMP algorithm consists of two steps: pre-selection and tree search. We first describe the pre-selection process and then discuss the efficient greedy tree search.
%

\subsection{Pre-selection: A First Stage Pruning}
\label{sec:preselection}

The purpose of the pre-selection is to estimate indices that are highly likely to be the elements of support $T$. Alternatively put, we do our best guess to choose columns of sensing matrix that are associated with nonzero elements of the sparse vector. Denoting the set of indices as $\Theta$, then the search set is reduced from $\Omega = \{1, 2,\cdots , N\}$ to $\Theta$, a small subset of $\Omega$.
When we perform the tree search, we only use elements of the pre-selected set $\Theta$ as a new element in the child paths so that we can limit the number of paths in the tree and eventually reduce the search complexity.
In the construction of $\Theta$, one can basically use any sparse recovery algorithm returning more than $K$ indices. Well-known examples include the OMP algorithm running more than K-iterations \cite{tong2011sparse} or the generalized OMP algorithm \cite{wang2012gomp}.
\subsection{Tree Search with Pruning}
\label{sec:treepruning}
%

%
Once the pre-selection is finished, we perform the tree search to identify the support.
In this setting, the tree has a maximum depth $K$, and the goal is to find a path with depth $K$ (i.e., candidate with cardinality $K$) that has the smallest cost function $J(\Lambda) = \| \mathbf{y} - \mathbf{\Phi}_\Lambda \hat{\mathbf{x}}_\Lambda \|_2$.
This cost function is often referred to as $\ell_2$-norm of the residual $\mathbf{r}_\Lambda = \mathbf{y} - \mathbf{\Phi}_\Lambda \hat{\mathbf{x}}_\Lambda$.
In each iteration, new child path is generated by adding new element to the existing path.
If we denote the path\footnote{In this paper, we use path and candidate interchangeably. In particular, we denote a full-blown path $\bar{s}_1^K$ by candidate.} at layer (iteration) $i$ as $\hat{s}_1^i$, then $\hat{s}_1^i = \{ s_1 , s_2 , \cdots s_i \}$ is the causal set chosen in the first $i$ iterations.
Since visiting all possible child nodes to find out the optimal solution is clearly prohibitive, we introduce an aggressive pruning strategy to remove unpromising paths from the tree.
This pruning decision is done by comparing the cost function of the path and the pruning threshold chosen by the smallest cost function of all paths visited.
%
%

%
\begin{figure}[t]
\begin{center}
		\includegraphics[width=145mm]{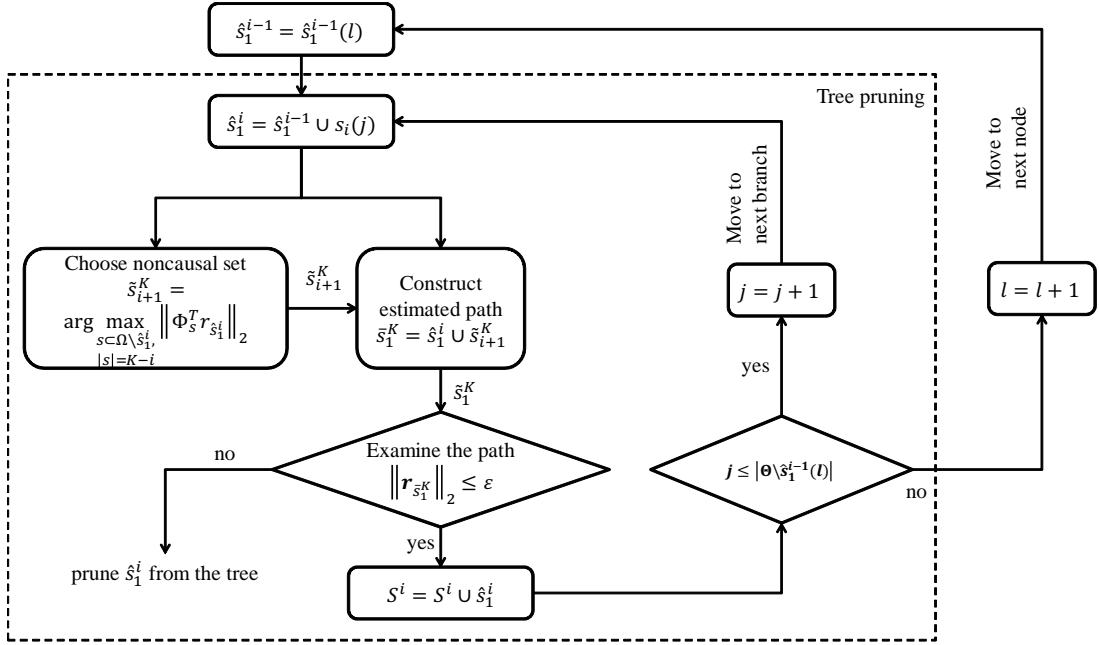}
		\caption{The pruning operation of TMP in the $i$-th layer where $j$ is the index of column in $\Theta \setminus \hat{s}_1^{i-1}$. Note that TMP investigates each path $\hat{s}_1^i$ and performs the pruning of the path if $\| \mathbf{r}_{\hat{s}_1^i \cup \tilde{s}_{i+1}^K} \|_2 > \epsilon$.}\label{fig:node_pruning}
\end{center}
\end{figure}
%

\begin{figure}
\begin{center}
		 \includegraphics[width=155mm]{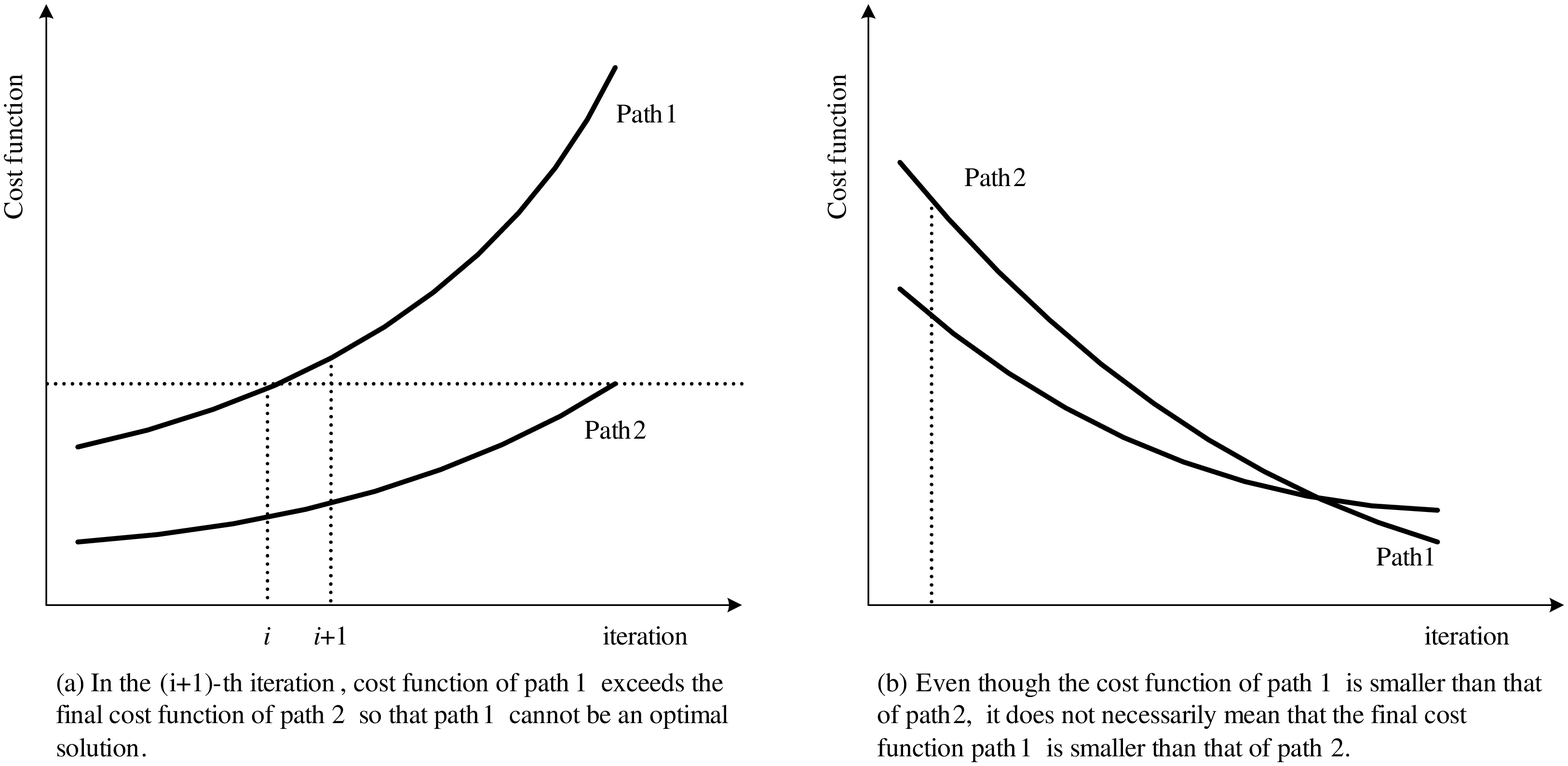} \caption{Cost function of the path: (a) conventional tree search and (b) proposed tree search.}\label{fig:costfunction}
\end{center}
\end{figure}

%
It is worth mentioning that in contrast to typical tree search problems, it is not easy, and in fact not possible, to decide the pruning of a path using the causal set only.
We note that in many tree search problems, the cost function of the path {\it increases monotonically} with the iteration (e.g., Viterbi decoding algorithm for maximum-likelihood detection) \cite{forney1973viterbi,viterbo1999universal,bshim2008sphere}.
Therefore, if a path whose partial cost function generated by the contributions of causal path only exceeds the cost function of already visited full-blown path, the path under investigation cannot be the solution of the problem and hence can be pruned immediately from the tree (see Fig. \ref{fig:costfunction}).
%
%
This pruning strategy, unfortunately, cannot be applied to the problem at hand since the partial cost function, which corresponds to the magnitude of the residual, is a {\it monotonic decreasing} function of the iteration\footnote{If $\hat{s}_1^i \subset \hat{s}_1^{i+1}$, then $\| \mathbf{r}_{\hat{s}_1^i} \|_2 \geq \|\mathbf{r}_{\hat{s}_1^{i+1}}\|_2$.}.
To make a proper decision, therefore, we have no way but to consider the cost function of full-blown path and hence need a noncausal set $\tilde{s}_{i+1}^K = \{ s_{i+1},\, s_{i+2},\, \cdots,\, s_K \}$ in the pruning process. 
This noncausal set $\tilde{s}_{i+1}^K$ is temporarily needed for the pruning operation and can be easily obtained by choosing $K- i$ indices of columns in $\Omega \setminus \hat{s}_1^{i}$ whose magnitude of the correlation with the residual $\mathbf{r}_{\hat{s}_1^i}$ is maximal\footnote{Instead of a single-shot process choosing $K-i$ indices simultaneously, noncausal set $\tilde{s}_{i+1}^K$ can be chosen by running multiple iterations for better judgement.}. That is,
%
\begin{eqnarray}
\label{eq:supp_est}
\tilde{s}_{i+1}^K = \arg \mathop{ \max_{s \subset \Omega \setminus \hat{s}_1^i,}} \limits_{|s| = K-i} \left\| \mathbf{\Phi}_{s}' \mathbf{r}_{\hat{s}_1^i} \right\|_2
\end{eqnarray}
%
where
%
\begin{eqnarray}
\mathbf{r}_{\hat{s}_1^i} &=& \mathbf{y} - \mathbf{\Phi}_{\hat{s}_1^i} \hat{\mathbf{x}}_{\hat{s}_1^i}, \nonumber \\
\hat{\mathbf{x}}_{\hat{s}_1^i} &=& \mathbf{\Phi}^{\dagger}_{\hat{s}_1^i} \mathbf{y}.\nonumber
\end{eqnarray}
%
For example, if $K - i = 2$, $\Omega \setminus \hat{s}_1^i = \{5, 7, 9, 11, \cdots \}$, and
$$|\mathbf{\phi}_{7}' \mathbf{r}_{\hat{s}_1^i} | > |\mathbf{\phi}_{11}' \mathbf{r}_{\hat{s}_1^i} | > | \mathbf{\phi}_{5}' \mathbf{r}_{\hat{s}_1^i} | > | \mathbf{\phi}_{9}' \mathbf{r}_{\hat{s}_1^i} | > \cdots,$$
then the noncausal set is $\tilde{s}_{i+1}^K = \{ 7, 11 \}$.

Once {\it roughly} estimated candidate $\bar{s}_1^K = \hat{s}_1^i \cup \tilde{s}_{i+1}^K$ is obtained, we compute the residual $\mathbf{r}_{\bar{s}_1^K} = \mathbf{y} - \mathbf{\Phi}_{\bar{s}_1^K} \hat{\mathbf{x}}_{\bar{s}_1^K}$ ($\hat{\mathbf{x}}_{\bar{s}_1^K} = \mathbf{\Phi}_{\bar{s}_1^K}^\dagger \mathbf{y}$) to decide whether to prune this path or not.
%
To be specific, if the $\ell_2$-norm of the residual is greater than the threshold $\epsilon$ (i.e., $\| \mathbf{r}_{\bar{s}_1^K} \|_2 > \epsilon$), then the path has little hope to survive and hence is pruned immediately (see Fig. \ref{fig:big_pic}).
Note that the pruning threshold $\epsilon$ is initialized to a large number and whenever the search of a layer is finished, updated to the minimum $\ell_2$-norm of the residual among all survived paths ($\epsilon = \min \| \mathbf{r}_{\bar{s}_1^K} \|_2$). Once the search is finished, a path with the minimum cost function is chosen as the final output of TMP.
We summarize the proposed TMP algorithm in Table \ref{tab:alg}.
%

\section{Performance Analysis}
\label{sec:analysis}

In this section, we analyze the recovery conditions under which TMP can accurately identify $K$-sparse signals in noiseless and noisy scenarios.
%
%
%
In our analysis, we use the restricted isometry property (RIP) of the sensing matrix.
%
\begin{definition}
\label{def_rip}
The sensing matrix $\mathbf{\Phi}$ is said to satisfy the RIP of order $K$ if there exists a constant $\delta(\mathbf{\Phi}) \in (0,1)$ such that
%
\begin{equation}
	(1-\delta(\mathbf{\Phi})) \| \mathbf{x} \|_2^2 \leq \| \mathbf{\Phi}\mathbf{x} \|_2^2 \leq (1+\delta(\mathbf{\Phi})) \| \mathbf{x} \|_2^2	\nonumber \label{eq:rip}
\end{equation}
%
for any $K$-sparse vector $\mathbf{x}$.
\end{definition}
%
In particular, the minimum of all constants $\delta (\mathbf{\Phi})$ satisfying Definition \ref{def_rip} is called the restricted isometry constant (RIC) and denoted by $\delta_K (\mathbf{\Phi})$. In the sequel, we use $\delta_K$ instead of $\delta_K (\mathbf{\Phi})$ for brevity.

In our analysis, we use the generalized OMP (gOMP) as a pre-selection algorithm. The gOMP algorithm chooses $L$ ($> 1$) indices in each iteration and hence $LK$ indices are chosen in total. Due to the selection of multiple indices, more than one true indices (indices in the support) can be chosen in each iteration and the chance of identifying the support increases substantially \cite{wang2012gomp}.
%

The following lemmas are useful in our analysis.
%
\begin{lemma}{\it (Lemma 3 in \cite{candes2005decoding}): }
\label{lem_monotone_rip}
If the sensing matrix $\mathbf{\Phi}$ satisfies the RIP of both orders $K_1$ and $K_2$, then $\delta_{K_1} < \delta_{K_2}$ for any $K_1 < K_2$.
\end{lemma}
%
%
\begin{lemma}{\it (Consequences of RIP \cite{candes2005decoding,needell2009cosamp}): }
\label{lem_consqc_rip}
If $0 < \delta_{|I|} < 1$ exists for $I \subset \Omega$, then for any vector $\mathbf{x} \in \mathbb{R}^{|I|}$,
%
\begin{eqnarray}
\label{eq:lem_consqc_rip}
\left( 1 - \delta_{|I|} \right) \| \mathbf{x} \|_2 \leq \| \mathbf{\Phi}_{I}' \mathbf{\Phi}_{I} \mathbf{x} \|_2 \leq \left( 1 + \delta_{|I|} \right) \| \mathbf{x} \|_2, \nonumber \\
\frac{1}{1+\delta_{|I|}} \| \mathbf{x} \|_2 \leq \| \left( \mathbf{\Phi}_{I}' \mathbf{\Phi}_{I} \right)^{-1} \mathbf{x} \|_2 \leq \frac{1}{1-\delta_{|I|}} \| \mathbf{x} \|_2 \nonumber.
\end{eqnarray}
%
\end{lemma}
%
%
\begin{lemma}{\it (Lemma 2.1 in \cite{candes2008restricted}): }
\label{lem_rip_disjoint_set}
Let $I_1,\, I_2 \subset \Omega$ and $I_1 \cap I_2 = \emptyset$. If $0 < \delta_{|I_1| + |I_2|} < 1$ exists, then
%
\begin{eqnarray}
\label{eq:lem_rip_disjoint_set}
\| \mathbf{\Phi}_{I_1}' \mathbf{\Phi}_{I_2} \mathbf{x} \|_2 \leq \delta_{|I_1| + |I_2|} \| \mathbf{x} \|_2. \nonumber
\end{eqnarray}
%
\end{lemma}
%

\subsection{Recovery from Noiseless Measurements}
\label{sec:noiseless}
%
%
In this subsection, we analyze a condition ensuring that TMP recovers the original sparse signal accurately from the noiseless measurements.
As mentioned, TMP consists of {\it pre-selection} and {\it tree search}. 
%
%
%
In our analysis, we show that the recovery condition of TMP is not much different from the condition of the pre-selection only and in fact guaranteed under more relaxed RIP bound (see Theorem \ref{thm_noiseless_suff_cond1}).
%
%

In order to ensure the accurate identification of the support, TMP should satisfy the following two conditions:
%
\begin{enumerate}
\item At least one support index should be selected in the pre-selection process (i.e., $T \cap \Theta \neq \emptyset$).
\item At least one true path\footnote{If $\hat{s}_1^i$ is a true path, it contains indices only in $T$ ($\hat{s}_1^i \subset T$).} should be survived in the tree pruning process.
\end{enumerate}
%
The following Theorem describes the condition ensuring that at least one support is identified by the pre-selection stage.
%
\begin{theorem}[Recovery condition in first iteration for noiseless scenario \cite{wang2012gomp}]
\label{thm_gomp_erc}
The gOMP algorithm identifies at least one support index in the first iteration if the sensing matrix $\mathbf{\Phi}$ satisfies
%
\begin{equation}
\label{eq:thm_gomp}
\delta_{L+K} < \frac{\sqrt{L}}{\sqrt{L}+\sqrt{K}}.
\end{equation}
%
\end{theorem}
%
\vspace{0.3cm}

We next analyze the condition ensuring that the final candidate $\bar{s}_1^K$ of the tree search equals the support $T$. In order to guarantee $\bar{s}_1^K = T$, at least one true path should be survived in each layer and further a true index should be added to this path.

Before we proceed, we provide definitions useful in our analysis.
Let $\lambda_i$ be the smallest correlation in magnitude between the residual $\mathbf{r}_{\hat{s}_1^i}$ and columns associated with {\it correct} indices. 
That is,
$$\lambda^i = \mathop{\min} \limits_{u \in T \setminus \hat{s}_1^i} \left| \phi_u' \mathbf{r}_{\hat{s}_1^i} \right|.$$
Further, let $\gamma^i$ be the largest correlation in magnitude between $\mathbf{r}_{\hat{s}_1^i}$ and columns associated with {\it incorrect} indices. 
That is,
$$\gamma^i = \mathop{\max}\limits_{u \in T^c} \left| < \phi_u , \mathbf{r}_{\hat{s}_1^i} > \right|.$$

In the following lemmas, we provide a lower bound of $\lambda^i$ and an upper bound of $\gamma^i$.
%
%
%
%
\begin{lemma}\label{lem:low_bound_a}
Suppose a path $\hat{s}_1^i$ is contained in $T$ (i.e., $\hat{s}_1^i \subset T$), then
\begin{equation}
\lambda^i \geq \frac{1 - \delta_K - \delta_M}{1-\delta_K} \left\| \mathbf{x}_{T \setminus \hat{s}_1^i } \right\|_2.	 \label{eq:low_bound_a}
\end{equation}
\begin{IEEEproof}
See Appendix \ref{app:lem:low_bound_a}.
\end{IEEEproof}
\end{lemma}

%
%
%
%
\begin{lemma}\label{lem:up_bound_b}
Suppose a path $\hat{s}_1^i$ is contained in $T$, then
\begin{equation}
\gamma^i \leq \frac{\delta_{K+1}}{1-\delta_K} \left\| \mathbf{x}_{T\setminus \hat{s}_1^i} \right\|_2.	 \label{eq:up_bound_b}
\end{equation}
\begin{IEEEproof}
See Appendix \ref{app:lem:up_bound_b}.
\end{IEEEproof}
\end{lemma}

As mentioned, in order to recover the original sparse signals, at least one true path should be survived in each layer. In other words, when a path $\hat{s}_1^i$ is contained in $T$ ($\hat{s}_{1}^i \subset T$), then the noncausal set $\tilde{s}_{i+1}^K$ should also be contained in $T$ (i.e., $\tilde{s}_{i+1}^K \subset T$) and further this path should not be pruned for the accurate reconstruction of the sparse signals.
%
%
That is,
%
\begin{equation}
\| \mathbf{r}_{\hat{s}_1^i \cup \tilde{s}_{i+1}^K} \| = \| \mathbf{r}_T \| < \epsilon.
\label{eq:pruning}
\end{equation}
%
Since $\| \mathbf{r}_T \|_2 = 0$ for the noiseless scenario, the condition \eqref{eq:pruning} always holds for any positive $\epsilon$. Thus, what we essentially need is a condition ensuring that the noncausal set chosen from \eqref{eq:supp_est} is contained in $T$ (i.e., $\tilde{s}_{i+1}^K \subset T$).
%
%
%
%
\begin{figure}[t]
\begin{center}
		\includegraphics[width=140mm]{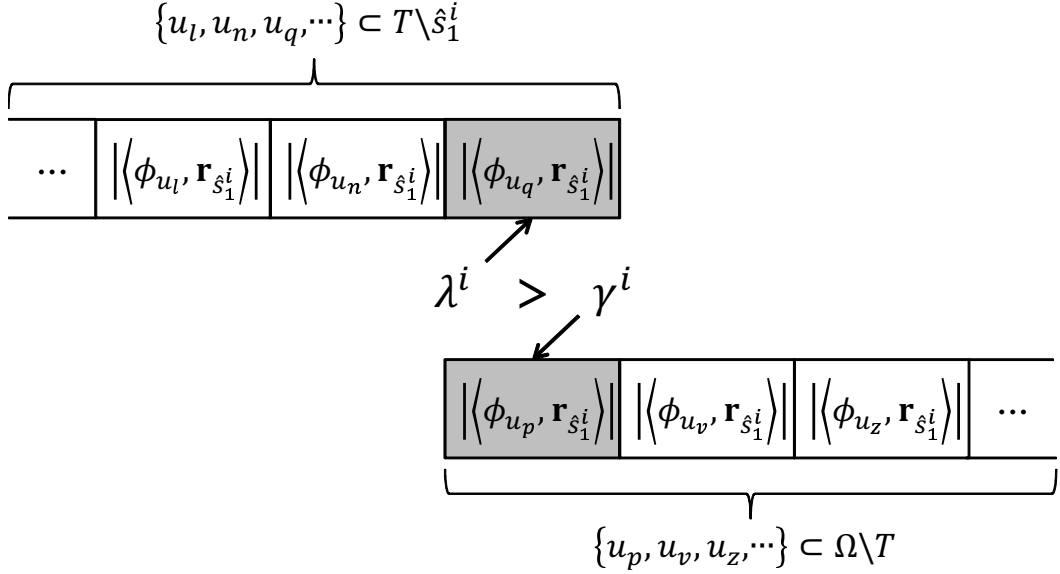}
		\caption{Comparison between $\lambda^i$ and $\gamma^i$ for the true path $\hat{s}_1^i$ (i.e., $\hat{s}_1^i \subset T$). $\lambda_i$ is the smallest correlation (in magnitude) between the residual $\mathbf{r}_{\hat{s}_1^i}$ and columns associated with $T \setminus \hat{s}_1^i$ and $\gamma^i$ is the largest correlation (in magnitude) between $\mathbf{r}_{\hat{s}_1^i}$ and columns associated with $T^c$. One can observe that, if $\lambda^i > \gamma^i$, then the noncausal set $\bar{s}_1^K$ is contained in $T$ ($\bar{s}_1^K = T \setminus \hat{s}_1^i$).}\label{fig:noiseless_erc}
\end{center}
\end{figure}
%
%
\begin{theorem}
\label{thm_noiseless2}
If the causal path $\hat{s}_1^i$ contains correct indices only, then the noncausal set $\tilde{s}_{i+1}^K$ of TMP consist only of correct ones under
%
\begin{equation}
\label{eq:thm_noiseless2}
\delta_M < \frac{1}{3}.
\end{equation}
%
for any $i$ ($0 \leq i \leq K - 1$). In other words, $\bar{s}_{1}^K (= \hat{s}_1^i \cup \tilde{s}_{i+1}^K) = T$ under $\delta_M < \frac{1}{3}$.
\begin{IEEEproof}
Since the indices of columns highly correlated with $\mathbf{r}_{\hat{s}_1^i}$ are chosen as elements of $\tilde{s}_{i+1}^K$ (see \eqref{eq:supp_est}), if $\lambda^i$ is larger than $\gamma^i$, then the noncausal set $\tilde{s}_{i+1}^K$ is contained in $T$ ($\tilde{s}_{i+1}^K = T \setminus \hat{s}_1^i$). In other words, $\bar{s}_1^K = T$ under
\begin{equation}
\lambda^i > \gamma^i.	\label{eq:ab_con}
\end{equation}
%
Using Lemma \ref{lem:low_bound_a} and \ref{lem:up_bound_b}, \eqref{eq:ab_con} holds under
$$\frac{1 - \delta_K - \delta_M}{1-\delta_K} \left\| \mathbf{x}_{T \setminus \hat{s}_1^i } \right\|_2 > \frac{\delta_{K+1}}{1-\delta_K} \left\| \mathbf{x}_{T\setminus \hat{s}_1^i} \right\|_2,$$
and hence $\delta_K + \delta_{K+1} + \delta_M < 1$.
Further, using Lemma \ref{lem_monotone_rip}, we have $3 \delta_M < 1$, which is the desired result.
\end{IEEEproof}
\end{theorem}

If Theorem \ref{thm_gomp_erc} and \ref{thm_noiseless2} are jointly satisfied, $\tilde{s}_{i+1}^K \subset T$ for any true path $\hat{s}_1^i$ ($\hat{s}_1^i \subset T$) so that $\bar{s}_1^K = \hat{s}_1^i \cup \tilde{s}_{i+1}^K = T$ and thus $\hat{s}_1^i$ will not be pruned from the tree (we recall that $\| \mathbf{r}_{\bar{s}_1^K} \|_2 = 0 < \epsilon$ for any positive $\epsilon$).
%
%
%
%
Therefore, overall recovery condition of TMP for the noiseless scenario can be obtained by combining Theorem \ref{thm_gomp_erc} and \ref{thm_noiseless2}.
%
\begin{theorem}[Recovery condition of TMP]
\label{thm_noiseless_suff_cond1}
The TMP algorithm identifies the support of any $K$-sparse signal from $\mathbf{y} = \mathbf{\Phi x}$ accurately if the sensing matrix $\mathbf{\Phi}$ satisfies the RIP with
%
\begin{eqnarray}
\delta_Q < \frac{1}{3} & & \text{if } K < 4L, \label{eq:thm_noiseless_suff_cond1} \\
\delta_Q < \frac{\sqrt{L}}{\sqrt{L}+\sqrt{K}} & & \text{otherwise} \label{eq:thm_noiseless_suff_cond12}
\end{eqnarray}
%
where $Q = \max\{M,L+K\}$.
%
\begin{IEEEproof}
The conditions \eqref{eq:thm_noiseless_suff_cond1} and \eqref{eq:thm_noiseless_suff_cond12} are obtained by choosing stricter condition between Theorem \ref{thm_gomp_erc} and \ref{thm_noiseless2}.
Specifically,
%
\begin{eqnarray}
\text{if } \frac{\sqrt{L}}{\sqrt{L}+\sqrt{K}} > \frac{1}{3} &\rightarrow& \delta_Q < \frac{1}{3} \nonumber \\
\text{otherwise } &\rightarrow& \delta_Q < \frac{\sqrt{L}}{\sqrt{L}+\sqrt{K}}, \nonumber
\end{eqnarray}
%
which is the desired result.
%
\end{IEEEproof}
\end{theorem}
%
\vspace{0.3cm}
Recall that the exact recovery condition of the original gOMP algorithm is \cite{wang2012gomp}
%
\begin{eqnarray}
\label{eq:gomp_cond}
\delta_{LK} < \frac{\sqrt{L}}{\sqrt{L} + 2\sqrt{K}}.
\end{eqnarray}
%
From \eqref{eq:thm_noiseless_suff_cond1}-\eqref{eq:gomp_cond}, it is clear that TMP provides more relaxed upper bound for any $L < K$ since $\frac{\sqrt{L}}{\sqrt{L} + 2\sqrt{K}} < \min\{ \frac{1}{3},\,\frac{\sqrt{L}}{\sqrt{L}+\sqrt{K}} \}$. Even if $L=K$, TMP is effective since the exact recovery conditions of gOMP and TMP are $$\delta_{K^2} < \frac{1}{3}$$ and $$\delta_{\max\{M, 2K\}} < \frac{1}{3},$$ respectively.
One can observe that in the large dimensional system satisfying $K^2 > M$, recovery condition of TMP is better (more relaxed) than the condition of gOMP. Similar argument holds for other sparse recovery algorithm (e.g., CoSaMP \cite{needell2009cosamp}).
%

%
\subsection{Reconstruction from Noisy Measurements}
\label{sec:noiseless}
%

Now, we turn to the noisy scenario and analyze the condition of TMP to accurately identify the support in the presence of noise.
Even though the details are a bit cumbersome, main architecture of the proof is reminiscent of the argument in the noiseless scenario.
In fact, two requirements of TMP to identify the support are 1) at least one support element should be chosen in the pre-selection process (i.e., $T \cap \Omega \neq \emptyset$), and 2) true path ($\bar{s}_1^K = T$) should be survived in the pruning process.
%
%
%

%
Before we proceed, we provide useful definitions in our analysis.
%
%
First, let $\rho$ be the largest correlation (in magnitude) between the observation $\mathbf{y}$ and the columns associated with true indices. That is,
%
\begin{eqnarray}
\label{eq:rho_def}
\rho = \max_{j \in T} |\phi_j' \mathbf{y}|. \nonumber
\end{eqnarray}
%
Next, let $\eta$ be the $L$-th largest correlation (in magnitude) between the observation $\mathbf{y}$ and the columns associated with incorrect indices. Then $\eta$ is expressed as
%
\begin{eqnarray}
\label{eq:eta_def}
\eta = \min_{j \in I_L} | \phi_j' \mathbf{y} |. \nonumber
\end{eqnarray}
%
where $I_L = \arg \mathop{\max} \limits_{|I|=L, I \subset T^c} \| \mathbf{\Phi}_I' \mathbf{r}_\Lambda \|_2$.

In the following lemmas, we provide the lower bound of $\rho$ and the upper bound of $\eta$.
%
\begin{lemma}
\label{lem_rho_lb}
$\rho$ satisfies
%
\begin{eqnarray}
\label{eq:lem_rho_lb}
\rho \geq \frac{1}{\sqrt{K}} \left[ \left( 1 - \delta_{K} \right) \| \mathbf{x}_{T} \|_2 - \sqrt{1+\delta_{K}} \| \mathbf{v} \|_2 \right]
\end{eqnarray}
%

\begin{IEEEproof}
See Appendex \ref{app:rho_lb}.
\end{IEEEproof}
\end{lemma}
%

%
\begin{lemma}
\label{lem_eta_ub}
$\eta$ satisfies
%
\begin{eqnarray}
\label{eq:lem_eta_ub}
\eta \leq \frac{1}{ \sqrt{L} } \left[ \delta_{L+K} \|\mathbf{x}_T \|_2 + \sqrt{1 + \delta_L} \| \mathbf{v} \|_2 \right].
\end{eqnarray}
%

\begin{IEEEproof}
See Appendex \ref{app:eta_ub}.
\end{IEEEproof}
\end{lemma}
%

The following theorem provides the condition ensuring that at least one support element is identified by the pre-selection stage.
%
\begin{theorem}
\label{lem_noisy_success_pre_selection}
The gOMP algorithm identifies at least one support element if the nonzero coefficients of the original sparse signal $\mathbf{x}$ satisfy
%
\begin{eqnarray}
\label{eq:thm_noisy_success_pre_selection}
\min_{j \in T} |x_j| > \frac{ (\sqrt{K} + \sqrt{L}) \sqrt{1 + \delta_{L+K}} }{ \sqrt{L} (1 - \delta_{K}) - \sqrt{K} \delta_{L+K} } \| \mathbf{v} \|_2.
\end{eqnarray}
%
%

\begin{IEEEproof}
From definitions of $\rho$ and $\eta$, it is clear that gOMP selects at least one true index in the first iteration if
%
%
\begin{eqnarray}
\label{eq:pf_noisy_success_pre_selection1}
\rho > \eta.
\end{eqnarray}
%
%
Using the lower bound of $\rho$ and the upper bound of $\eta$, we obtain the sufficient condition of \eqref{eq:pf_noisy_success_pre_selection1} as
%
\begin{eqnarray}
\label{eq:pf_noisy_success_pre_selection12}
\frac{1}{\sqrt{K}} \left[ \left( 1 - \delta_{K} \right) \| \mathbf{x}_{T} \|_2 - \sqrt{1+\delta_{K}} \| \mathbf{v} \|_2 \right] > \frac{1}{ \sqrt{L} } \left[ \delta_{L+K} \|\mathbf{x}_T \|_2 + \sqrt{1 + \delta_L} \| \mathbf{v} \|_2 \right].
\end{eqnarray}
%
After some manipulations, we have
%
\begin{eqnarray}
\label{eq:pf_noisy_success_pre_selection2}
\| \mathbf{x}_T \|_2 > \frac{ (\sqrt{K} + \sqrt{L}) \sqrt{1 + \delta_{L+K}}  }{ \sqrt{L} (1 - \delta_{K}) - \sqrt{K} \delta_{L+K} }\| \mathbf{v} \|_2.
\end{eqnarray}
%
Since $\| \mathbf{x}_T \|_2 \geq \mathop{\min} \limits_{j \in T} |x_j|$, \eqref{eq:pf_noisy_success_pre_selection2} is guaranteed under
%
\begin{eqnarray}
\label{eq:pf_noisy_success_pre_selection3}
\min_{j \in T} |x_j| > \frac{ (\sqrt{K} + \sqrt{L}) \sqrt{1 + \delta_{L+K}} }{ \sqrt{L} (1 - \delta_{K}) - \sqrt{K} \delta_{L+K} }\| \mathbf{v} \|_2,
\end{eqnarray}
%
which completes the proof.
\end{IEEEproof}

\end{theorem}


\vspace{0.1cm}
We next analyze the condition under which the true path is survived by the tree pruning stage.
In order to meet this requirement, 1) under the condition that a causal set is true ($\hat{s}_1^i \subset T$), corresponding noncausal set should also be true ($\tilde{s}_{i+1}^K \subset T$) and further 2) this true path should not be removed by the tree pruning (i.e., if $\bar{s}_1^K = T$, then $\| \mathbf{r}_{\bar{s}_1^K} \|_2 < \epsilon$).

Before we proceed, we introduce two useful definitions in our analysis. Let $\beta^i$ be the smallest correlation in magnitude between $\phi_j$ ($j \in T \setminus \hat{s}_1^i$) and $\mathbf{r}_{\hat{s}_1^i}$:
%
\begin{eqnarray}
\label{eq:beta_def}
\beta^i = \arg \min_{j \in T \setminus \hat{s}_1^i} |\phi_j' \mathbf{r}_{\hat{s}_1^i}|. \nonumber
\end{eqnarray}
%
Similarly, let $\alpha^i$ be the largest correlation in magnitude between $\phi_j$ ($j \in T^c$) and the residual $\mathbf{r}_{\hat{s}_1^i}$:
%
\begin{eqnarray}
\label{eq:alpha_def}
\alpha^i = \arg \max_{j \in T^c} |\phi_j' \mathbf{r}_{\hat{s}_1^i}|. \nonumber
\end{eqnarray}
%
The following two lemmas provide the lower and upper bounds of $\beta^i$ and $\alpha^i$, respectively.
%
\begin{lemma}
\label{lem_beta_lb}
If $\hat{s}_1^i$ contains true indices exclusively, then $\beta^i$ satisfies
%
\begin{eqnarray}
\label{eq:lem_beta_lb}
\beta^i \geq \left( 1 - \delta_M - \frac{ \delta_{K+1} \delta_K }{ 1 - \delta_K } \right) \| \mathbf{x}_{T \setminus \hat{s}_1^i} \|_2 - \sqrt{1+\delta_M} \| \mathbf{v} \|_2
\end{eqnarray}
%
\begin{IEEEproof}
See Appendix \ref{app:beta_lb}.
\end{IEEEproof}
\end{lemma}
%
%
\begin{lemma}
\label{lem_alpha_ub}
If $\hat{s}_1^i$ contains true indices exclusively, then $\alpha^i$ satisfies
%
\begin{eqnarray}
\label{eq:lem_alpha_ub}
\alpha^i \leq \left( \delta_{K+1} + \frac{ \delta_{K+1} \delta_K }{ 1 - \delta_K } \right) \| \mathbf{x}_{T \setminus \hat{s}_1^i} \|_2 + \sqrt{1 + \delta_M} \| \mathbf{v} \|_2
\end{eqnarray}
%
\begin{IEEEproof}
See Appendix \ref{app:alpha_ub}.
\end{IEEEproof}
\end{lemma}
%
%
Using these lemmas, we can identify the condition guaranteeing that the noncausal set $\tilde{s}_{i+1}^K$ of a true path ($\hat{s}_1^i \subset T$) is also true.
%
\begin{lemma}
\label{lem_noisy_success_find_T}
Suppose a causal path $\hat{s}_1^i$ consists of true indices exclusively (i.e., $\hat{s}_1^i \subset T$), then the noncausal set $\tilde{s}_{i+1}^K$ also contains the true ones ($\tilde{s}_{i+1}^K = T \setminus \hat{s}_1^i$) under
%
\begin{eqnarray}
\label{eq:lem_noisy_find_T}
\min_{j \in T} |x_j| > \frac{2(1-\delta_K)\sqrt{1+\delta_M}}{1-\delta_K-\delta_{K+1}-\delta_M}\| \mathbf{v} \|_2.
\end{eqnarray}
%

\begin{IEEEproof}
One can easily show that the noncausal set of any true path $\hat{s}_1^i$ contains only true indices if
%
\begin{eqnarray}
\label{eq:pf_noisy_success_find_T1}
\beta^i > \alpha^i.
\end{eqnarray}
%
Using Lemma \ref{lem_beta_lb} and \ref{lem_alpha_ub}, we obtain the sufficient condition of \eqref{eq:pf_noisy_success_find_T1} as
\begin{eqnarray}
\label{eq:pf_noisy_find_T3}
\lefteqn{ \left( 1 - \delta_M - \frac{ \delta_M \delta_K }{ 1 - \delta_K } \right) \| \mathbf{x}_{T \setminus \hat{s}_1^i} \|_2 - \sqrt{1+\delta_M} \| \mathbf{v} \|_2 } \nonumber \\
& & \hspace{3cm} > \left( \delta_{K+1} + \frac{ \delta_{K+1} \delta_K }{ 1 - \delta_K } \right) \| \mathbf{x}_{T \setminus \hat{s}_1^i} \|_2 + \sqrt{1 + \delta_M} \| \mathbf{v} \|_2.
\end{eqnarray}
%
After some manipulations, we have
%
\begin{eqnarray}
\label{eq:lem_noisy_find_T4}
\| \mathbf{x}_{T \setminus \hat{s}_1^i} \|_2 > \frac{2(1-\delta_K)\sqrt{1+\delta_M}}{1-\delta_K-\delta_{K+1}-\delta_M}\|\mathbf{v}\|_2.
\end{eqnarray}
%
Since $\| \mathbf{x}_{T \setminus \hat{s}_1^i} \|_2 \geq \mathop{\min} \limits_{j \in T} |x_j|$, we get the desired result.
\end{IEEEproof}

\end{lemma}
%
Next, we turn to the analysis of the condition under which the magnitude of $\mathbf{r}_T$ becomes the minimum among all combinations of $K$ indices.
%
\begin{lemma}
\label{lem_noisy1}
The candidate whose residual is minimum (in magnitude) becomes the support if
%
\begin{equation}
\min_{j \in T} |x_j| > \frac{ 2 (1 - \delta_K) }{ 1 - 3 \delta_{2K} }\|\mathbf{v}\|_2.
\label{eq:sc_gen_n}
\end{equation}
%
In other words, $\| \mathbf{r}_T \|_2 < \| \mathbf{r}_{\bar{s}_1^K} \|_2$ for any $\bar{s}_1^K \neq T$ under \eqref{eq:sc_gen_n}.
%
\begin{IEEEproof}
One can notice that the hypothesis is satisfied if the upper bound of $\| \mathbf{r}_T \|_2$ is smaller than the lower bound of $\| \mathbf{r}_{\bar{s}_1^K} \|_2$. First, we obtain the upper bound of $\| \mathbf{r}_T \|_2$ as
%
\begin{eqnarray}
\| \mathbf{r}_T \|_2
&=& \| \mathbf{P}_T^\bot \mathbf{y} \|_2 \label{eq:pf_sc_gen_n11} \nonumber\\
&=& \| \mathbf{P}_T^\bot \left( \mathbf{\Phi}_T \mathbf{x}_T + \mathbf{v} \right) \|_2 \label{eq:pf_sc_gen_n12} \nonumber\\
&=& \| \mathbf{P}_T^\bot \mathbf{v} \|_2 \label{eq:pf_sc_gen_n13} \nonumber\\
&\leq& \| \mathbf{v} \|_2 \label{eq:pf_sc_gen_n1}
\end{eqnarray}
%
where $\mathbf{P}_T^\bot = \mathbf{I} - \mathbf{\Phi}_T \left( \mathbf{\Phi}_T'\mathbf{\Phi}_T \right)^{-1} \mathbf{\Phi}_T'$ is the projection onto the orthogonal complement of $T$ and \eqref{eq:pf_sc_gen_n13} is because $\mathbf{P}_T^\bot \mathbf{\Phi}_T \mathbf{x}_T = \mathbf{0}$.

Next, we obtain the lower bound of $\| \mathbf{r}_{\bar{s}_1^K} \|_2$. For any $\bar{s}_1^K \neq T$, we have
%
\begin{eqnarray}
\label{eq:pf_sc_gen_n2}
\| \mathbf{r}_{\bar{s}_1^K} \|_2
&=& \| \mathbf{P}_{\bar{s}_1^K}^\bot \mathbf{y} \|_2 \label{eq:pf_sc_gen_n213} \nonumber \\
&=& \| \mathbf{P}_{\bar{s}_1^K}^\bot ( \mathbf{\Phi x} + \mathbf{v} ) \|_2 \label{eq:pf_sc_gen_n212} \nonumber \\
&=& \| \mathbf{P}_{\bar{s}_1^K}^\bot ( \mathbf{\Phi}_{\bar{s}_1^K} \mathbf{x}_{\bar{s}_1^K} + \mathbf{\Phi}_{T \setminus \bar{s}_1^K} \mathbf{x}_{T \setminus \bar{s}_1^K} + \mathbf{v} ) \|_2 \label{eq:pf_sc_gen_n211} \nonumber \\
&=& \| \mathbf{P}_{\bar{s}_1^K}^\bot ( \mathbf{\Phi}_{T \setminus \bar{s}_1^K} \mathbf{x}_{T \setminus \bar{s}_1^K} + \mathbf{v} ) \|_2 \label{eq:pf_sc_gen_n21} \\
&\geq& \| \mathbf{P}_{\bar{s}_1^K}^\bot \mathbf{\Phi}_{T \setminus \bar{s}_1^K} \mathbf{x}_{T \setminus \bar{s}_1^K} \|_2 - \| \mathbf{P}_{\bar{s}_1^K}
^\bot \mathbf{v} \|_2 \label{eq:pf_sc_gen_n22}
\end{eqnarray}
%
where the inequality in \eqref{eq:pf_sc_gen_n21} is because $\mathbf{P}_{\bar{s}_1^K}^\bot \mathbf{\Phi}_{\bar{s}_1^K}^\bot \mathbf{x}_{\bar{s}_1^K}^\bot = \mathbf{0}$ and the inequality in \eqref{eq:pf_sc_gen_n22} is due to the triangle inequality.
The first term in the right-hand side of \eqref{eq:pf_sc_gen_n22} is lower bounded as
%
\begin{eqnarray}
\label{eq:pf_sc_gen_n3}
\| \mathbf{P}_{\bar{s}_1^K}^\bot \mathbf{\Phi}_{T \setminus \bar{s}_1^K} \mathbf{x}_{T \setminus \bar{s}_1^K} \|_2
&=& \| (\mathbf{I} - \mathbf{\Phi}_{\bar{s}_1^K} ( \mathbf{\Phi}_{\bar{s}_1^K}'\mathbf{\Phi}_{\bar{s}_1^K} )^{-1} \mathbf{\Phi}_{\bar{s}_1^K}') \mathbf{\Phi}_{T \setminus \bar{s}_1^K} \mathbf{x}_{T \setminus \bar{s}_1^K} \|_2 \nonumber \\
&\geq& \| \mathbf{\Phi}_{T \setminus \bar{s}_1^K} \mathbf{x}_{T \setminus \bar{s}_1^K} \|_2 - \| \mathbf{\Phi}_{\bar{s}_1^K} ( \mathbf{\Phi}_{\bar{s}_1^K}'\mathbf{\Phi}_{\bar{s}_1^K} )^{-1} \mathbf{\Phi}_{\bar{s}_1^K}' \mathbf{\Phi}_{T \setminus \bar{s}_1^K} \mathbf{x}_{T \setminus \bar{s}_1^K} \|_2 \\
&\geq& \lefteqn{ \sqrt{1 - \delta_{|T \setminus \bar{s}_1^K|}} \| \mathbf{x}_{T \setminus \bar{s}_1^K} \|_2 } \nonumber \\
& & - \sqrt{1 + \delta_{|\bar{s}_1^K|}} \| ( \mathbf{\Phi}_{\bar{s}_1^K}'\mathbf{\Phi}_{\bar{s}_1^K} )^{-1} \mathbf{\Phi}_{\bar{s}_1^K}' \mathbf{\Phi}_{T \setminus \bar{s}_1^K} \mathbf{x}_{T \setminus \bar{s}_1^K} \|_2 \label{eq:pf_sc_gen_n31} \\
&\geq& \sqrt{1 - \delta_{|T \setminus \bar{s}_1^K|}} \| \mathbf{x}_{T \setminus \bar{s}_1^K} \|_2 - \frac{ \sqrt{1 + \delta_{|\bar{s}_1^K|}} }{ 1 - \delta_{|\bar{s}_1^K|} } \| \mathbf{\Phi}_{\bar{s}_1^K}' \mathbf{\Phi}_{T \setminus \bar{s}_1^K} \mathbf{x}_{T \setminus \bar{s}_1^K} \|_2 \label{eq:pf_sc_gen_n32} \\
&\geq& \sqrt{1 - \delta_{|T \setminus \bar{s}_1^K|}} \| \mathbf{x}_{T \setminus \bar{s}_1^K} \|_2 - \frac{ \sqrt{1 + \delta_{|\bar{s}_1^K|}} \delta_{K + |T \setminus \bar{s}_1^K|} }{ 1 - \delta_{|\bar{s}_1^K|} } \| \mathbf{x}_{T \setminus \bar{s}_1^K} \|_2 \label{eq:pf_sc_gen_n33} \\
&>& \sqrt{1 - \delta_{2K} } \| \mathbf{x}_{T \setminus \bar{s}_1^K} \|_2 - \frac{ \sqrt{1 + \delta_K} \delta_{2K} }{ 1 - \delta_K} \| \mathbf{x}_{T \setminus \bar{s}_1^K} \|_2 \label{eq:pf_sc_gen_n34}
\end{eqnarray}
%
where \eqref{eq:pf_sc_gen_n31} is from Definition \ref{def_rip}, \eqref{eq:pf_sc_gen_n32} is from from Lemma \ref{lem_consqc_rip}, and \eqref{eq:pf_sc_gen_n33} and \eqref{eq:pf_sc_gen_n34} are from Lemma \ref{lem_rip_disjoint_set} and \ref{lem_monotone_rip}, respectively.
Using this together with $\|\mathbf{P}_{\bar{s}_1^K}^\bot \mathbf{v}\|_2 \leq \| \mathbf{v} \|_2$, we have
%
\begin{equation}
\label{eq:pf_sc_gen_n36}
\| \mathbf{r}_{\bar{s}_1^K} \|_2 > \sqrt{1 - \delta_{2K} } \| \mathbf{x}_{T \setminus \bar{s}_1^K} \|_2 - \frac{ \sqrt{1 + \delta_K} \delta_{2K} }{ 1 - \delta_K} \| \mathbf{x}_{T \setminus \bar{s}_1^K} \|_2 - \| \mathbf{v} \|_2.
\end{equation}
%
for any $\bar{s}_1^K \neq T$.
Since $\| \mathbf{r}_T \|_2 < \| \mathbf{r}_{\bar{s}_1^K} \|_2$ always holds if the upper bound of $\| \mathbf{r}_T \|_2$ is smaller than the lower bound of $\| \mathbf{r}_{\bar{s}_1^K} \|_2$, it is clear from \eqref{eq:pf_sc_gen_n1} and \eqref{eq:pf_sc_gen_n36} that the hypothesis is satisfied under
%
\begin{equation}
\label{eq:pf_sc_gen_n4}
\| \mathbf{x}_{T \setminus \bar{s}_1^K} \|_2 > \frac{ 2 (1 - \delta_K) }{ 1 - 3 \delta_{2K} }\|\mathbf{v}\|_2.
\end{equation}
%
Noting that $\| \mathbf{x}_{T \setminus \bar{s}_1^K} \|_2 \geq \mathop{\min} \limits_{j \in T} |x_j|$, we get the desired result.
\end{IEEEproof}
%
\end{lemma}
%

Thus far, we investigated the condition under which the noncausal set is true when the causal path is true (Lemma \ref{lem_noisy_success_find_T}) and the condition ensuring that the true path has the minimum residual (in magnitude) and hence survives during the tree pruning (Lemma \ref{lem_noisy1}).
%
Recalling that the pruning threshold is updated by the minimum value of the residual (in magnitude) in each layer ($\epsilon = \min \| \mathbf{r}_{\bar{s}_1^K} \|_2$) and a path whose residual magnitude is larger than $\epsilon$ is pruned, the support $T$ will never be pruned if the conditions of Lemma \ref{lem_noisy_success_find_T} and \ref{lem_noisy1} are jointly satisfied.
Formal description of our findings is as follows.
%
%
%
%
\begin{theorem}
\label{lem_noisy_T_survive}
The true path $\hat{s}_1^i \subset T$ survives in the pruning process for any $i$ under
%
\begin{eqnarray}
\label{eq:lem_noisy_T_survive}
\min_{j \in T} |x_j| > \max ( \mu, \omega )\|\mathbf{v}\|_2
\end{eqnarray}
%
where $\mu = \frac{2(1-\delta_K)}{1-3\delta_{2K}}$ and $\omega = \frac{2(1-\delta_K)\sqrt{1+\delta_M}}{1-\delta_K-\delta_{K+1}-\delta_M}$.
%
\begin{IEEEproof}
Immediate from Lemma \ref{lem_noisy_success_find_T} and \ref{lem_noisy1}.
\end{IEEEproof}
\end{theorem}

\vspace{0.15cm}

By combining the results of pre-selection (Theorem \ref{lem_noisy_success_pre_selection}) and tree search (Theorem \ref{lem_noisy_T_survive}), we obtain the main result for the noisy setting.
%
\begin{theorem}
\label{thm_noisy_success}
The TMP algorithm accurately identifies the support from the noisy measurement $\mathbf{y} = \mathbf{\Phi}\mathbf{x} + \mathbf{v}$ under
%
\begin{eqnarray}
\label{eq:thm_noisy_success0}
\min_{j \in T} |x_j| > \gamma \|\mathbf{v}\|_2
\end{eqnarray}
%
$\gamma = \max ( \nu, \mu, \omega )$ and $\mu = \frac{2(1-\delta_K)}{1-3\delta_{2K}}$, $\omega = \frac{2(1-\delta_K)\sqrt{1+\delta_M}}{1-\delta_K-\delta_{K+1}-\delta_M}$, and $\nu = \frac{ (\sqrt{K} + \sqrt{L}) \sqrt{1 + \delta_{L+K}} \| \mathbf{v} \|_2 }{ \sqrt{L} (1 - \delta_{K}) - \sqrt{K} \delta_{L+K} }$.

\begin{IEEEproof}
Immediate from Theorem \ref{lem_noisy_success_pre_selection} and \ref{lem_noisy_T_survive}.
%
\end{IEEEproof}
\end{theorem}

It is worth noting that under \eqref{eq:thm_noisy_success0}, which essentially corresponds to the high signal-to-noise ratio (SNR) regime, we can identify the exact support information so that we can simply remove all non-support elements (zero entries in $\mathbf{x}$) and columns associated with these from the system model. In doing so, we can obtain the {\it overdetermined} system $\mathbf{y} = \mathbf{\Phi}_T \mathbf{x}_T + \mathbf{v}$ and the reconstructed signal becomes equivalent to the output of the best possible estimator referred to as Oracle estimator $\hat{\mathbf{x}} = \mathbf{\Phi}^{\dagger}_T \mathbf{y}$.
%
%
%
%
%

Using the part of analysis we obtained, we can also show the stability of the TMP algorithm.
By stability, we mean that the $\ell_2$-norm of the estimation error $\| \mathbf{x} - \hat{\mathbf{x}}_{\bar{s}_1^K} \|_2 = \| \mathbf{x} - \mathbf{\Phi}_{\bar{s}_1^K}^\dagger \mathbf{y} \|_2$ is upper bounded by the constant multiple of the noise power.
%
\begin{theorem}
\label{thm_noisy2}
The output $\hat{\mathbf{x}}_{\bar{s}_1^K}$ of the TMP algorithm satisfies
%
\begin{eqnarray}
\label{eq:thm_noisy2}
\left\| \mathbf{x} - \hat{\mathbf{x}}_{\bar{s}_1^K} \right\|_2 < \tau \| \mathbf{v} \|_2
\end{eqnarray}
where $\tau = \frac{ (\gamma + 1) (1 - \delta_K) + 2\gamma\delta_{2K} }{ (1 - \delta_K) \sqrt{1 - \delta_{2K}} }$.

\begin{IEEEproof}
From Definition \ref{def_rip}, it is clear that
%
\begin{eqnarray}
\label{eq:pf_noisy2_1}
\| \mathbf{x} - \hat{\mathbf{x}}_{\bar{s}_1^K} \|_2 \leq \frac{ \| \mathbf{\Phi} ( \mathbf{x} - \hat{ \mathbf{x} }_{\bar{s}_1^K} ) \|_2 }{ \sqrt{1 - \delta_{|T \cup \bar{s}_1^K|} }}.
\end{eqnarray}
%
Since $\mathbf{x} - \hat{\mathbf{x}}$ is at most $2K$-sparse, we further have
%
\begin{eqnarray}
\label{eq:pf_noisy2_2}
\| \mathbf{x} - \hat{\mathbf{x}}_{\bar{s}_1^K} \|_2
&\leq& \frac{ \| \mathbf{\Phi} ( \mathbf{x} - \hat{ \mathbf{x} }_{\bar{s}_1^K} ) \|_2 }{ \sqrt{1 - \delta_{|T \cup \bar{s}_1^K|} }} \nonumber \\
&=& \frac{ \| \mathbf{\Phi} ( \mathbf{x} - (\mathbf{\Phi}_{\bar{s}_1^K}' \mathbf{\Phi}_{\bar{s}_1^K})^{-1} \mathbf{\Phi}_{\bar{s}_1^K}' \mathbf{y} ) \|_2 }{ \sqrt{1 - \delta_{2K} }} \nonumber \\
&=& \frac{ \| \mathbf{\Phi} \mathbf{x} - \mathbf{\Phi}_{\bar{s}_1^K} (\mathbf{\Phi}_{\bar{s}_1^K}' \mathbf{\Phi}_{\bar{s}_1^K})^{-1} \mathbf{\Phi}_{\bar{s}_1^K}' ( \mathbf{\Phi x} + \mathbf{v} ) \|_2 }{ \sqrt{1 - \delta_{2K} }} \nonumber \\
&=& \frac{ \| \mathbf{P}_{\bar{s}_1^K}^\bot \mathbf{\Phi}_{T \setminus \bar{s}_1^K} \mathbf{x}_{T \setminus \bar{s}_1^K} - \mathbf{P}_{\bar{s}_1^K} \mathbf{v} \|_2 }{ \sqrt{1 - \delta_{2K} }} \nonumber \\
&\leq& \frac{ \| \mathbf{P}_{\bar{s}_1^K}^\bot \mathbf{\Phi}_{T \setminus \bar{s}_1^K} \mathbf{x}_{T \setminus \bar{s}_1^K} \|_2 + \| \mathbf{P}_{\bar{s}_1^K} \mathbf{v} \|_2 }{ \sqrt{1 - \delta_{2K} }} \nonumber \\
&\leq& \frac{ \| \mathbf{P}_{\bar{s}_1^K}^\bot \mathbf{\Phi}_{T \setminus \bar{s}_1^K} \mathbf{x}_{T \setminus \bar{s}_1^K} \|_2 + \| \mathbf{v} \|_2 }{ \sqrt{1 - \delta_{2K} }}
\end{eqnarray}
%
where $\mathbf{P}_{\bar{s}_1^K} = \mathbf{\Phi}_{\bar{s}_1^K} (\mathbf{\Phi}_{\bar{s}_1^K}' \mathbf{\Phi}_{\bar{s}_1^K})^{-1} \mathbf{\Phi}_{\bar{s}_1^K}'$.
Also,
%
\begin{eqnarray}
\label{eq:pf_noisy2_3}
\| \mathbf{P}_{\bar{s}_1^K}^\bot \mathbf{\Phi}_{T \setminus \bar{s}_1^K} \mathbf{x}_{T \setminus \bar{s}_1^K} \|_2
&=& \| (\mathbf{I} - \mathbf{\Phi}_{\bar{s}_1^K} ( \mathbf{\Phi}_{\bar{s}_1^K}'\mathbf{\Phi}_{\bar{s}_1^K} )^{-1} \mathbf{\Phi}_{\bar{s}_1^K}') \mathbf{\Phi}_{T \setminus \bar{s}_1^K} \mathbf{x}_{T \setminus \bar{s}_1^K} \|_2 \nonumber \\
&\leq& \| \mathbf{\Phi}_{T \setminus \bar{s}_1^K} \mathbf{x}_{T \setminus \bar{s}_1^K} \|_2 + \| \mathbf{\Phi}_{\bar{s}_1^K} ( \mathbf{\Phi}_{\bar{s}_1^K}'\mathbf{\Phi}_{\bar{s}_1^K} )^{-1} \mathbf{\Phi}_{\bar{s}_1^K}' \mathbf{\Phi}_{T \setminus \bar{s}_1^K} \mathbf{x}_{T \setminus \bar{s}_1^K} \|_2 \\
&\leq& \lefteqn{ \sqrt{1 + \delta_{|T \setminus \bar{s}_1^K|}} \| \mathbf{x}_{T \setminus \bar{s}_1^K} \|_2 } \nonumber \\
& & + \sqrt{1 + \delta_{|\bar{s}_1^K|}} \| ( \mathbf{\Phi}_{\bar{s}_1^K}'\mathbf{\Phi}_{\bar{s}_1^K} )^{-1} \mathbf{\Phi}_{\bar{s}_1^K}' \mathbf{\Phi}_{T \setminus \bar{s}_1^K} \mathbf{x}_{T \setminus \bar{s}_1^K} \|_2 \label{eq:pf_noisy2_31} \\
&\leq& \sqrt{1 + \delta_{|T \setminus \bar{s}_1^K|}} \| \mathbf{x}_{T \setminus \bar{s}_1^K} \|_2 + \frac{ \sqrt{1 + \delta_{|\bar{s}_1^K|}} }{ 1 - \delta_{|\bar{s}_1^K|} } \| \mathbf{\Phi}_{\bar{s}_1^K}' \mathbf{\Phi}_{T \setminus \bar{s}_1^K} \mathbf{x}_{T \setminus \bar{s}_1^K} \|_2 \label{eq:pf_noisy2_32} \\
&\leq& \sqrt{1 + \delta_{|T \setminus \bar{s}_1^K|}} \| \mathbf{x}_{T \setminus \bar{s}_1^K} \|_2 + \frac{ \sqrt{1 + \delta_{|\bar{s}_1^K|}} \delta_{K + |T \setminus \bar{s}_1^K|} }{ 1 - \delta_{|\bar{s}_1^K|} } \| \mathbf{x}_{T \setminus \bar{s}_1^K} \|_2 \label{eq:pf_noisy2_33} \\
&\leq& \sqrt{1 + \delta_{K} } \| \mathbf{x}_{T \setminus \bar{s}_1^K} \|_2 + \frac{ \sqrt{1 + \delta_K} \delta_{2K} }{ 1 - \delta_K} \| \mathbf{x}_{T \setminus \bar{s}_1^K} \|_2 \label{eq:pf_noisy2_34} \\
&<& \sqrt{1 + \delta_{2K} } \| \mathbf{x}_{T \setminus \bar{s}_1^K} \|_2 + \frac{ \sqrt{1 + \delta_K} \delta_{2K} }{ 1 - \delta_K} \| \mathbf{x}_{T \setminus \bar{s}_1^K} \|_2 \\
&<& \frac{ 1 -\delta_K + 2\delta_{2K} }{ 1 - \delta_K } \| \mathbf{x}_{T \setminus \bar{s}_1^K} \|_2 \label{eq:pf_noisy2_35}
\end{eqnarray}
%
where \eqref{eq:pf_noisy2_31} is from Definition \ref{def_rip}, and \eqref{eq:pf_noisy2_32} and \eqref{eq:pf_noisy2_33} are from Lemma \ref{lem_consqc_rip} and \ref{lem_rip_disjoint_set}, respectively.
Plugging \eqref{eq:pf_noisy2_35} into \eqref{eq:pf_noisy2_2}, we have
%
\begin{eqnarray}
\label{eq:pf_noisy2_4}
\| \mathbf{x} - \hat{\mathbf{x}}_{\bar{s}_1^K} \|_2
< \frac{ ( 1 - \delta_K + 2\delta_{2K} ) \| \mathbf{x}_{T \setminus \bar{s}_1^K} \|_2 }{ (1 - \delta_K) \sqrt{1 - \delta_{2K}} } + \frac{ \| \mathbf{v} \|_2 }{ \sqrt{1 - \delta_{2K} }}.
\end{eqnarray}
%
Note that when the support is chosen accurately, $\bar{s}_1^K = T$ and thus
%
\begin{eqnarray}
\label{eq:pf_noisy2_4_1}
\| \mathbf{x}_{T \setminus \bar{s}_1^K} \|_2 = 0.
\end{eqnarray}
%
Whereas, if $\bar{s}_1^K \neq T$, then by the contraposition of Theorem \ref{thm_noisy_success}\footnote{Here, we need to use slightly modified version of Theorem \ref{thm_noisy_success}, which says that if $\| \mathbf{x}_{\bar{s}_1^K} \|_2 > \gamma \| \mathbf{v} \|_2$, then $\bar{s}_1^K = T$.}, we have
%
\begin{eqnarray}
\label{eq:pf_noisy2_5}
\| \mathbf{x}_{T \setminus \bar{s}_1^K} \|_2
\leq \gamma \| \mathbf{v} \|_2.
\end{eqnarray}
%
for any $\bar{s}_1^K \neq T$.
By combining \eqref{eq:pf_noisy2_4_1} and \eqref{eq:pf_noisy2_5}, we obtain the desired result.
\end{IEEEproof}

\end{theorem}
\section{Simulation and Discussions}
\label{sec:simulation}

\subsection{Simulation Setup}
\label{sec:simulation_setup}

In this section, we observe the performance of sparse recovery algorithms including TMP through empirical simulations.
In our simulations, we generate $K$-sparse vector $\mathbf{x}$ whose nonzero locations and coefficients are randomly chosen and the sensing matrix $\mathbf{\Phi}$ of size $100 \times 256$ whose entries are from the independent Gaussian distribution $\mathcal{N}(0, \frac{1}{M})$.
In each point of the individual recovery algorithm, we perform at least $n = 5,000$ independent trials.
%
%
In the noiseless setting, we use the exact recovery ratio (ERR) as a performance measure. In the noisy setting, we use the mean squared error (MSE) of the recovery algorithms which is defined as
$$MSE = \frac{1}{n} \mathop{\sum} \limits_{\ell=1}^n \frac{\| \mathbf{x}(\ell) - \hat{\mathbf{x}}(\ell) \|^2}{N}$$
where $\hat{\mathbf{x}}(\ell)$ is the estimate of the original sparse signal $\mathbf{x}(\ell)$.
%
%

We test simulations on the following algorithms:
%
\begin{enumerate}
\item OMP algorithm \cite{caiomp}
\item BP algorithm \cite{tibshirani1996}: we use BP in noiseless setting and basis pursuit denoising (BPDN) in noisy setting.
\item CoSaMP algorithm \cite{needell2009cosamp}: we set the maximal number of iterations to $40$.
\item gOMP algorithm \cite{wang2012gomp}: we choose two indices ($L=2$) in each iteration.
\item TMP: we use gOMP ($L=2$) in the pre-selection stage.
\item TMP with limited branching: we set the maximum number of branches in each layer ($N_{\max} = 10$ and $100$).
\end{enumerate}
%

\subsection{Simulation Results}
\label{sec:simulation_result}
%

%
\begin{figure}
\begin{center}
\includegraphics[width=140mm, height=125mm]{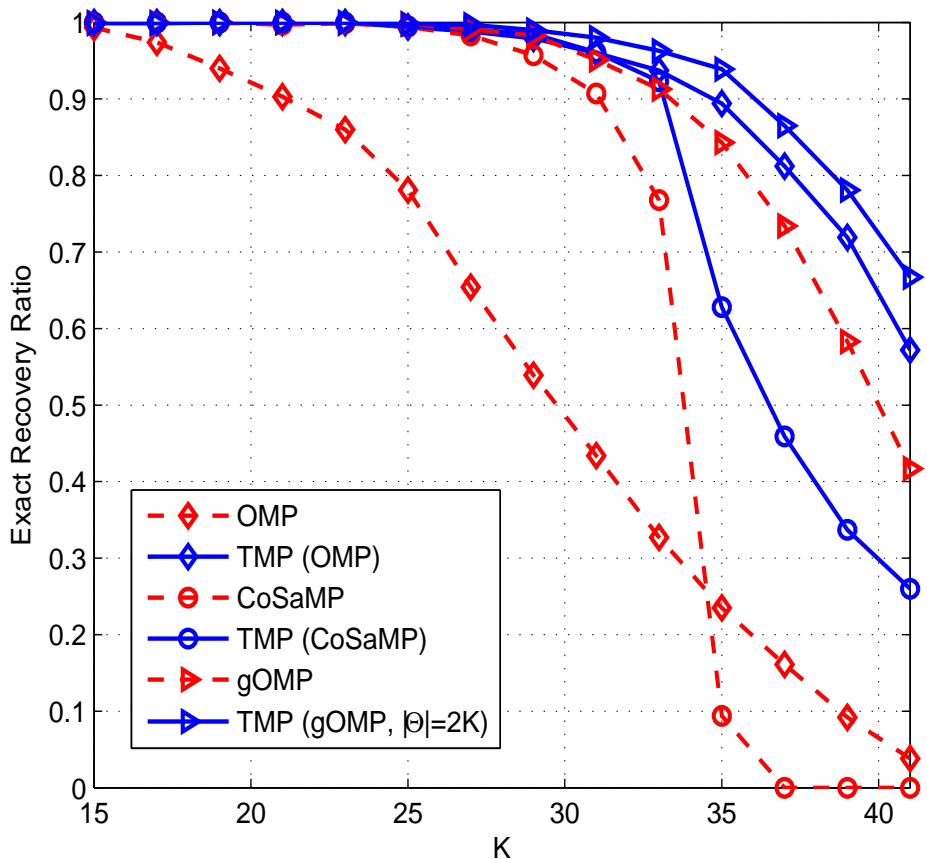} \caption{ERR performance as a function of the sparsity $K$ in the noiseless setting. We measure the performance before the tree search (pre-selection) and after the tree search (TMP).}\label{fig:err}
\end{center}
\end{figure}
%
We first compare the ERR performance of sparse recovery algorithms in the noiseless setting. Main purpose of this simulation is to observe how much performance gain can be achieved by the tree search.
%
Since we use the conventional sparse recovery algorithm in the pre-selection process, effectiveness of the proposed TMP algorithm can be checked by comparing the recovery performance before the tree search (pre-selection only) and after the tree search.
%
%
\begin{figure}
\begin{center}
		 \includegraphics[width=140mm, height=125mm]{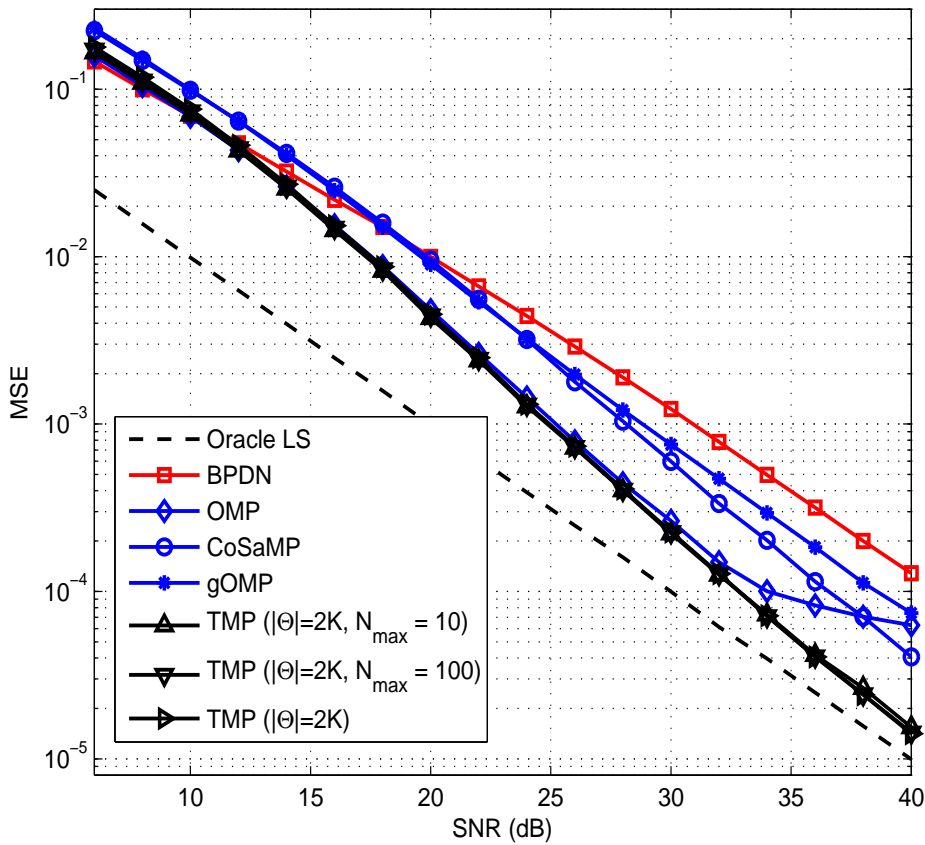}
		\caption{MSE performance of sparse recovery algorithms ($K = 20$) in $100 \times 256$ system.}\label{fig:mse20}
\end{center}
\end{figure}
%
In Fig. \ref{fig:err}, we plot the ERR of TMP with various pre-selection algorithms as a function of the sparsity $K$. Overall, we observe that the addition of tree search process provides substantial gain in performance.
In particular, when $K$ is large, performance gain obtained by the tree search stage is noticeable.
When OMP is used as a pre-selection algorithm, for example, the ERR of TMP before and after the tree search at $K = 35$ are $0.23$ and $0.89$, respectively.
%
%
%
%

In Fig. \ref{fig:mse20}, we plot the MSE performance of the sparse recovery algorithms as a function of signal-to-noise ratio (SNR) in the noisy setting. Note that the decibel (dB) scale of SNR is defined as $\mbox{SNR} = 10 \log_{10} \frac{ \| \mathbf{\Phi x} \|^2 }{ \| \mathbf{v} \|^2 }$.
In this test, we set the sparsity level to $K=20$ so that $8\%$ of entries in the input vector are nonzero.
Overall, we observe that the performance gain of TMP improves with SNR. While the performance gap between the conventional sparse recovery algorithms and Oracle estimator is maintained across the board, the performance gap between TMP and Oracle estimator gets smaller as SNR increases.
%
\begin{figure}
\begin{center}
		 \includegraphics[width=140mm, height=125mm]{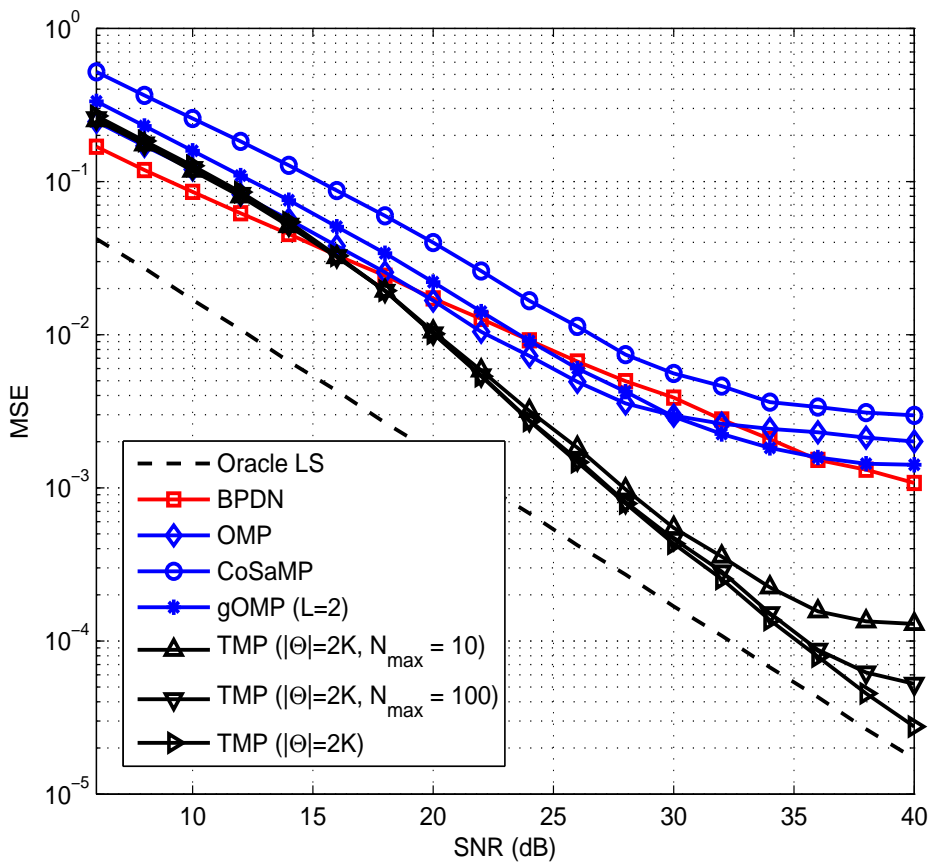}
		\caption{MSE performance of sparse recovery algorithms in the noisy setting ($K=30$).}\label{fig:mse30}
\end{center}
\end{figure}
%
In Fig. \ref{fig:mse30}, a similar simulation as before but with large $K$ is performed.
In this simulation, we set $K=30$ so that $12\%$ of entries are nonzero.
In this case, we clearly see that TMP outperforms conventional sparse recovery algorithms and the performance gain improves with SNR.
For example, the gain at $MSE=10^{-2}$ is around $2$ dB but the gain at $MSE=10^{-3}$ is more than $10$ dB.
Also, as it can be seen from the figure and also in accordance with Theorem \ref{thm_noisy_success}, the performance of TMP is asymptotically optimal in high SNR regime in the sense that it approaches the MSE performance of Oracle estimator.
%
%
%

%
\begin{figure}
\begin{center}
		 \includegraphics[width=140mm, height=125mm]{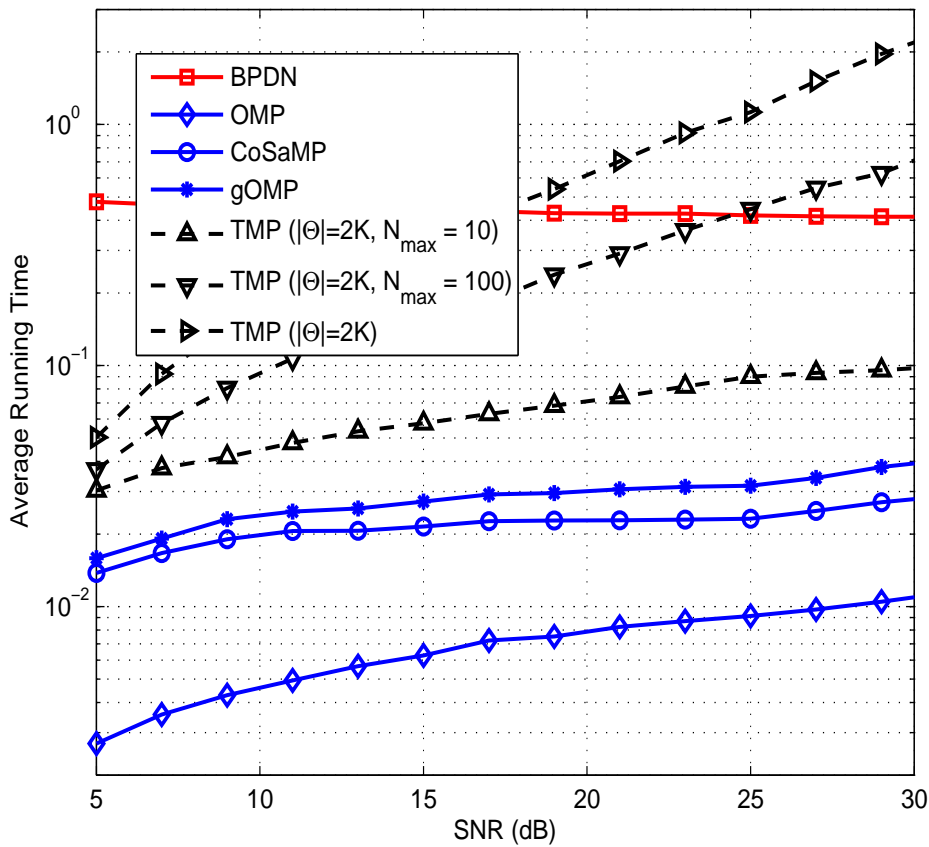}
		\caption{Average running time of sparse signal recovery algorithms in $100 \times 256$ system.}\label{fig:complexity}
\end{center}
\end{figure}
%
Fig. \ref{fig:complexity} shows the running time complexity of the sparse recovery algorithms as a function of the sparsity level $K$. All algorithms under test are coded by MATLAB software package and run by a personal computer with Intel Core i$5$ processor and Microsoft Windows $7$ environment.
As seen in the figure, among greedy algorithms under test, OMP exhibits the smallest running time. Since TMP performs tree search to investigate multiple promising paths, it is no wonder that the running time complexity of TMP is higher than the rest of greedy algorithms.
However, by limiting the number of branching operations, computational burden of TMP can be reduced dramatically. Due to the reduction in number of investigated paths, we can observe that the running time complexity of TMP with limited branching is much smaller than that without limitation. In particular, if $N_{\max} =10$, TMP achieves two order of magnitude reduction over the original TMP algorithm with only slight loss in performance.
\section{Conclusions}
\label{sec:conclusion}
%


In this paper, we proposed a tree search based sparse signal recovery algorithm referred to as matching pursuit with a tree pruning (TMP).
In order to overcome the shortcoming of greedy algorithm in choosing {\it short-sighted} candidates, the TMP algorithm performs the tree search and investigates multiple promising candidates.
The complexity overhead caused by the tree search is controlled by the pre-selection and tree pruning.
In our empirical simulation, we observed that TMP provides excellent recovery performance in both noiseless and noisy scenarios.
While TMP is promising algorithm in terms of the recovery performance, its complexity is a bit higher than existing greedy algorithms and further study is needed.
Our future work will address the complexity reduction issue of greedy tree search algorithm to achieve better tradeoff between complexity and performance.

\appendices

\section{Proof of Lemma \ref{lem:low_bound_a}}
\label{app:lem:low_bound_a}

Let $\hat{s}_1^i \subset T$ and $\lambda^i = \mathop{\min} \limits_{j \in T \setminus \hat{s}_1^i} |\phi_j' \mathbf{r}_{\hat{s}_1^i}|$, then by using the triangular inequality, we have
%
\begin{eqnarray}
\label{eq:app_lambda_lb1}
|\phi_j' \mathbf{r}_{\hat{s}_1^i}|
&=& \| \phi_j' \mathbf{r}_{\hat{s}_1^i} \|_2 \\
&=& \| \phi_j' \mathbf{P}_{\hat{s}_1^i}^\bot \mathbf{y} \|_2 \\
&=& \| \phi_j' \mathbf{P}_{\hat{s}_1^i}^\bot \mathbf{\Phi}_{T \setminus \hat{s}_1^i} \mathbf{x}_{T \setminus \hat{s}_1^i} \|_2 \label{eq:app_lambda_lb11} \\
&\geq& \| \phi_j' \mathbf{\Phi}_{T \setminus \hat{s}_1^i} \mathbf{x}_{T \setminus \hat{s}_1^i} \|_2 - \| \phi_j' \mathbf{P}_{\hat{s}_1^i} \mathbf{\Phi}_{T \setminus \hat{s}_1^i} \mathbf{x}_{T \setminus \hat{s}_1^i} \|_2 \label{eq:app_lambda_lb12}
\end{eqnarray}
%
where $\mathbf{P}_{\hat{s}_1^i}^\bot = \mathbf{I} - \mathbf{\Phi}_{\hat{s}_1^i} (\mathbf{\Phi}_{\hat{s}_1^i}'\mathbf{\Phi}_{\hat{s}_1^i})^{-1} \mathbf{\Phi}_{\hat{s}_1^i}'$.
%
From Definition \ref{def_rip}, we have
%
\begin{eqnarray}
\label{eq:app_lambda_lb2}
\| \phi_j' \mathbf{\Phi}_{T \setminus \hat{s}_1^i} \mathbf{x}_{T \setminus \hat{s}_1^i} \|_2
&\geq& \sqrt{1-\delta_M} \| \mathbf{\Phi}_{T \setminus \hat{s}_1^i} \mathbf{x}_{T \setminus \hat{s}_1^i} \|_2 \label{eq:app_lambda_lb22} \\
&\geq& \sqrt{1-\delta_M} \sqrt{1-\delta_{|T \setminus \hat{s}_1^i|}} \| \mathbf{x}_{T \setminus \hat{s}_1^i} \|_2. \label{eq:app_lambda_lb21}
\end{eqnarray}
%
Also,
%
\begin{eqnarray}
\| \phi_j' \mathbf{P}_{\hat{s}_1^i} \mathbf{\Phi}_{T \setminus \hat{s}_1^i} \mathbf{x}_{T \setminus \hat{s}_1^i} \|_2
&=& \| \phi_j' \mathbf{\Phi}_{\hat{s}_1^i} (\mathbf{\Phi}_{\hat{s}_1^i}'\mathbf{\Phi}_{\hat{s}_1^i})^{-1} \mathbf{\Phi}_{\hat{s}_1^i}' \mathbf{\Phi}_{T \setminus \hat{s}_1^i} \mathbf{x}_{T \setminus \hat{s}_1^i} \|_2 \\
&\geq& \delta_{|\hat{s}_1^i|+1} \| (\mathbf{\Phi}_{\hat{s}_1^i}'\mathbf{\Phi}_{\hat{s}_1^i})^{-1} \mathbf{\Phi}_{\hat{s}_1^i}' \mathbf{\Phi}_{T \setminus \hat{s}_1^i} \mathbf{x}_{T \setminus \hat{s}_1^i} \|_2 \label{eq:app_lambda_lb222} \\
&\geq& \frac{ \delta_{|\hat{s}_1^i|+1} }{ 1-\delta_{|\hat{s}_1^i|} } \| \mathbf{\Phi}_{\hat{s}_1^i}' \mathbf{\Phi}_{T \setminus \hat{s}_1^i} \mathbf{x}_{T \setminus \hat{s}_1^i} \|_2 \label{eq:app_lambda_lb23} \\
&\geq& \frac{ \delta_{|\hat{s}_1^i|+1} \delta_K }{ 1-\delta_{|\hat{s}_1^i|} } \| \mathbf{x}_{T \setminus \hat{s}_1^i} \|_2 \label{eq:app_lambda_lb24}
\end{eqnarray}
%
where \eqref{eq:app_lambda_lb21} is from Definition \ref{def_rip}, \eqref{eq:app_lambda_lb22} and \eqref{eq:app_lambda_lb24} are from Lemma \ref{lem_rip_disjoint_set}, and \eqref{eq:app_lambda_lb23} is from Lemma \ref{lem_consqc_rip}.
Using \eqref{eq:app_lambda_lb12}, \eqref{eq:app_lambda_lb21}, and \eqref{eq:app_lambda_lb24}, we have
%
\begin{eqnarray}
\label{eq:app_lambda_lb3}
|\phi_j' \mathbf{r}_{\hat{s}_1^i}|
&\geq& \| \phi_j' \mathbf{\Phi}_{T \setminus \hat{s}_1^i} \mathbf{x}_{T \setminus \hat{s}_1^i} \|_2 - \| \phi_j' \mathbf{P}_{\hat{s}_1^i} \mathbf{\Phi}_{T \setminus \hat{s}_1^i} \mathbf{x}_{T \setminus \hat{s}_1^i} \|_2 \nonumber \\
&\geq& \sqrt{1-\delta_M} \sqrt{1-\delta_{|T \setminus \hat{s}_1^i|}} \| \mathbf{x}_{T \setminus \hat{s}_1^i} \|_2 - \frac{ \delta_{|\hat{s}_1^i|+1} \delta_K }{ 1-\delta_{|\hat{s}_1^i|} } \| \mathbf{x}_{T \setminus \hat{s}_1^i} \|_2.
\end{eqnarray}
%
Since \eqref{eq:app_lambda_lb3} holds for any $j \in T \setminus \hat{s}_1^i$ and $0 \leq i \leq K$, we have
%
\begin{eqnarray}
\label{eq:app_lambda_lb4}
\lambda^i
&\geq& \sqrt{1-\delta_M} \sqrt{1-\delta_{|T \setminus \hat{s}_1^i|}} \| \mathbf{x}_{T \setminus \hat{s}_1^i} \|_2 - \frac{ \delta_{|\hat{s}_1^i|+1} \delta_K }{ 1-\delta_{|\hat{s}_1^i|} } \| \mathbf{x}_{T \setminus \hat{s}_1^i} \|_2 \\
&\geq& \sqrt{1-\delta_M} \sqrt{1-\delta_M} \| \mathbf{x}_{T \setminus \hat{s}_1^i} \|_2 - \frac{ \delta_{K+1} \delta_K }{ 1-\delta_K } \| \mathbf{x}_{T \setminus \hat{s}_1^i} \|_2 \\
&\geq& \left( 1-\delta_M - \frac{ \delta_M \delta_K }{ 1-\delta_K } \right) \| \mathbf{x}_{T \setminus \hat{s}_1^i} \|_2 \\
&=& \frac{ 1- \delta_K - \delta_M }{ 1-\delta_K } \| \mathbf{x}_{T \setminus \hat{s}_1^i} \|_2,
\end{eqnarray}
%
which is the desired result.

\section{Proof of Lemma \ref{lem:up_bound_b}}
\label{app:lem:up_bound_b}

Let $\hat{s}_1^i \subset T$ and $\gamma^i = \mathop{\max} \limits_{j \in T^c} |\phi_j' \mathbf{r}_{\hat{s}_1^i}|$, then by using the triangle inequality, we have
%
\begin{eqnarray}
|\phi_j' \mathbf{r}_{\hat{s}_1^i}|
&=& \| \phi_j' \mathbf{r}_{\hat{s}_1^i} \|_2 \\
&=& \| \phi_j' \mathbf{P}_{\hat{s}_1^i}^\bot \mathbf{y} \|_2 \\
&=& \| \phi_j' \mathbf{P}_{\hat{s}_1^i}^\bot \mathbf{\Phi}_{T \setminus \hat{s}_1^i} \mathbf{x}_{T \setminus \hat{s}_1^i} \|_2 \\
&\leq& \| \phi_j' \mathbf{\Phi}_{T \setminus \hat{s}_1^i} \mathbf{x}_{T \setminus \hat{s}_1^i} \|_2 + \| \phi_j' \mathbf{P}_{\hat{s}_1^i} \mathbf{\Phi}_{T \setminus \hat{s}_1^i} \mathbf{x}_{T \setminus \hat{s}_1^i} \|_2. \label{eq:app_gamma_ub1}
\end{eqnarray}
%
Since $j \in T^c$, we have
%
\begin{eqnarray}
\label{eq:app_gamma_ub2}
\| \phi_j' \mathbf{\Phi}_{T \setminus \hat{s}_1^i} \mathbf{x}_{T \setminus \hat{s}_1^i} \|_2
&\leq& \delta_{|T \setminus \hat{s}_1^i|+1} \| \mathbf{x}_{T \setminus \hat{s}_1^i} \|_2 \label{eq:app_gamma_ub21}
\end{eqnarray}
%
where \eqref{eq:app_gamma_ub21} is from Lemma \ref{lem_rip_disjoint_set}. Also,
%
\begin{eqnarray}
\| \phi_j' \mathbf{P}_{\hat{s}_1^i} \mathbf{\Phi}_{T \setminus \hat{s}_1^i} \mathbf{x}_{T \setminus \hat{s}_1^i} \|_2
&=& \| \phi_j' \mathbf{\Phi}_{\hat{s}_1^i} (\mathbf{\Phi}_{\hat{s}_1^i}'\mathbf{\Phi}_{\hat{s}_1^i})^{-1} \mathbf{\Phi}_{\hat{s}_1^i}' \mathbf{\Phi}_{T \setminus \hat{s}_1^i} \mathbf{x}_{T \setminus \hat{s}_1^i} \|_2 \label{eq:app_gamma_ub25} \\
&\leq& \delta_{|\hat{s}_1^i|+1} \| (\mathbf{\Phi}_{\hat{s}_1^i}'\mathbf{\Phi}_{\hat{s}_1^i})^{-1} \mathbf{\Phi}_{\hat{s}_1^i}' \mathbf{\Phi}_{T \setminus \hat{s}_1^i} \mathbf{x}_{T \setminus \hat{s}_1^i} \|_2 \label{eq:app_gamma_ub22} \\
&\leq& \frac{ \delta_{|\hat{s}_1^i|+1} }{ 1-\delta_{|\hat{s}_1^i|} } \| \mathbf{\Phi}_{\hat{s}_1^i}' \mathbf{\Phi}_{T \setminus \hat{s}_1^i} \mathbf{x}_{T \setminus \hat{s}_1^i} \|_2 \label{eq:app_gamma_ub23} \\
&\leq& \frac{ \delta_{|\hat{s}_1^i|+1} \delta_K }{ 1-\delta_{|\hat{s}_1^i|} } \| \mathbf{x}_{T \setminus \hat{s}_1^i} \|_2 \label{eq:app_gamma_ub24}
\end{eqnarray}
%
where \eqref{eq:app_gamma_ub22} and \eqref{eq:app_gamma_ub24} are from Lemma \ref{lem_rip_disjoint_set} and \eqref{eq:app_gamma_ub23} is from Lemma \ref{lem_consqc_rip}.
Using \eqref{eq:app_gamma_ub1}, \eqref{eq:app_gamma_ub21}, and \eqref{eq:app_gamma_ub24}, we have
%
\begin{eqnarray}
\label{eq:app_gamma_ub3}
|\phi_j' \mathbf{r}_{\hat{s}_1^i}|
&\leq& \| \phi_j' \mathbf{\Phi}_{T \setminus \hat{s}_1^i} \mathbf{x}_{T \setminus \hat{s}_1^i} \|_2 + \| \phi_j' \mathbf{P}_{\hat{s}_1^i} \mathbf{\Phi}_{T \setminus \hat{s}_1^i} \mathbf{x}_{T \setminus \hat{s}_1^i} \|_2 \\
&\leq& \left( \delta_{|T \setminus \hat{s}_1^i|+1} + \frac{ \delta_{|\hat{s}_1^i|+1} \delta_K }{ 1-\delta_{|\hat{s}_1^i|} } \right) \| \mathbf{x}_{T \setminus \hat{s}_1^i} \|_2 \\
&\leq& \left( \delta_{K+1} + \frac{ \delta_{K+1} \delta_K }{ 1-\delta_K } \right) \| \mathbf{x}_{T \setminus \hat{s}_1^i} \|_2 \label{eq:app_gamma_ub31} \\
&=& \frac{ \delta_{K+1} }{ 1-\delta_K } \| \mathbf{x}_{T \setminus \hat{s}_1^i} \|_2,
\end{eqnarray}
%
which is the desired result.
%

\section{Proof of Lemma \ref{lem_rho_lb}}
\label{app:rho_lb}

From the definition of $\rho$ in \eqref{eq:rho_def}, we have
%
\begin{eqnarray}
\rho
&=& \max_{j \in T} |\phi_j' \mathbf{y}| \label{eq:app_rho_lb0} \\
&=& \| \mathbf{\Phi}_{T}' \mathbf{y} \|_\infty \label{eq:app_rho_lb00} \\
&\geq& \frac{1}{ \sqrt{| T |} } \| \mathbf{\Phi}_{T}' \mathbf{y} \|_2 \label{eq:app_rho_lb1} \\
&=& \frac{1}{ \sqrt{K} } \| \mathbf{\Phi}_{T}' (\mathbf{\Phi}_{T} \mathbf{x}_{T} + \mathbf{v}) \|_2 \\
&\geq& \frac{1}{ \sqrt{K} } \left( \| \mathbf{\Phi}_{T}' \mathbf{\Phi}_T \mathbf{x}_{T} \|_2 - \| \mathbf{\Phi}_{T}' \mathbf{v} \|_2 \right) \! \label{eq:app_rho_lb000}
\end{eqnarray}
%
where \eqref{eq:app_rho_lb1} is from the inequality $\|\mathbf{u}\|_\infty \geq \frac{1}{\sqrt{\|\mathbf{u}\|_0}} \|\mathbf{u}\|_2$ for any vector $\mathbf{u}$.
Note that
%
\begin{eqnarray}
\| \mathbf{\Phi}_{T}' \mathbf{\Phi}_T \mathbf{x}_{T} \|_2
\geq (1 - \delta_{K}) \| \mathbf{x}_{T} \|_2 \label{eq:app_rho_lb1_1}
\end{eqnarray}
%
and
%
\begin{eqnarray}
\| \mathbf{\Phi}_{T}' \mathbf{v} \|_2
\leq \sqrt{1 + \delta_K} \| \mathbf{v} \|_2 \label{eq:app_rho_lb1_2}
\end{eqnarray}
%
and thus $\rho$ is lower bounded as
%
\begin{eqnarray}
\label{eq:app_rho_lb3}
\rho
\geq \frac{1}{ \sqrt{K} } \left[ (1 - \delta_{K}) \| \mathbf{x}_{T} \|_2 - \sqrt{1 + \delta_K} \| \mathbf{v} \|_2 \right],
\end{eqnarray}
%
which is the desired result.
%
\section{Proof of Lemma \ref{lem_eta_ub}}
\label{app:eta_ub}

From the definition of $\eta$ in \eqref{eq:eta_def}, we have
%
\begin{eqnarray}
\label{eq:app_eta_ub1}
\sqrt{ L } \eta \leq \sqrt{ \sum_{j \in I_L} | \phi_j' \mathbf{y}|^2 } = \| \mathbf{\Phi}_{I_L}' \mathbf{y} \|_2
\end{eqnarray}
%
where $I_L = \arg \mathop{\max} \limits_{|I|=L, I \subset T^c} \| \mathbf{\Phi}_I' \mathbf{y} \|_2$.
Using the triangle inequality, we have
%
\begin{eqnarray}
\| \mathbf{\Phi}_{I_L}' \mathbf{y} \|_2
&=& \| \mathbf{\Phi}_{I_L}' (\mathbf{\Phi}_T \mathbf{x}_T + \mathbf{v}) \|_2 \label{eq:app_eta_ub210} \\
&\leq& \| \mathbf{\Phi}_{I_L}' \mathbf{\Phi}_T \mathbf{x}_T \|_2 + \| \mathbf{\Phi}_{I_L}' \mathbf{v} \|_2. \label{eq:app_eta_ub2}
\end{eqnarray}
%
Since $I_L$ and $T$ are disjoint ($I_L \subset T^c$), we have
%
\begin{eqnarray}
\label{eq:app_eta_ub3}
\| \mathbf{\Phi}_{I_L}' \mathbf{\Phi}_T \mathbf{x}_T \|_2
\leq \delta_{L + K} \| \mathbf{x}_T \|_2
\end{eqnarray}
%
and
%
\begin{eqnarray}
\| \mathbf{\Phi}_{I_L}' \mathbf{v} \|_2
&\leq& \sqrt{1 + \delta_L} \| \mathbf{v} \|_2. \label{eq:app_eta_ub42}
\end{eqnarray}
%
Using \eqref{eq:app_eta_ub3} and \eqref{eq:app_eta_ub42}, we have
%
\begin{eqnarray}
\| \mathbf{\Phi}_{I_L}' \mathbf{y} \|_2
\leq \delta_{L+K} \| \mathbf{x}_T \|_2 + \sqrt{1+\delta_L} \| \mathbf{v} \|_2 \label{eq:app_eta_ub5}
\end{eqnarray}
%
and since $\| \mathbf{\Phi}_{I_L}' \mathbf{y} \|_2 \geq \sqrt{L} \eta$, we have
%
\begin{eqnarray}
\label{eq:app_eta_ub6}
\eta
\leq \frac{1}{ \sqrt{L} } \left[ \delta_{L+K} \| \mathbf{x}_T \|_2 + \sqrt{1+\delta_L} \| \mathbf{v} \|_2 \right],
\end{eqnarray}
%
which is the desired result.

%
\section{Proof of Lemma \ref{lem_beta_lb}}
\label{app:beta_lb}
%

Suppose $\hat{s}_1^i \subset T$ and $\beta^i = \mathop{\min} \limits_{j \in T \setminus \hat{s}_1^i} |\phi_j' \mathbf{r}_{\hat{s}_1^i}|$, then
%
\begin{eqnarray}
|\phi_j' \mathbf{r}_{\hat{s}_1^i}|
&=& \| \phi_j' \mathbf{r}_{\hat{s}_1^i} \|_2 \label{eq:app:beta_lb130} \\
&=& \| \phi_j' \mathbf{P}_{\hat{s}_1^i}^\bot \mathbf{y} \|_2 = \| \phi_j' \mathbf{P}_{\hat{s}_1^i}^\bot (\mathbf{\Phi}_T \mathbf{x}_T + \mathbf{v}) \|_2 \label{eq:app:beta_lb1300} \\
&=& \| \phi_j' \mathbf{P}_{\hat{s}_1^i}^\bot \mathbf{\Phi}_{T \setminus \hat{s}_1^i} \mathbf{x}_{T \setminus \hat{s}_1^i} + \phi_j' \mathbf{P}_{\hat{s}_1^i}^\bot \mathbf{v} \|_2 \label{eq:app:beta_lb13} \\
&=& \| \phi_j' \mathbf{\Phi}_{T \setminus \hat{s}_1^i} \mathbf{x}_{T \setminus \hat{s}_1^i} - \phi_j' \mathbf{P}_{\hat{s}_1^i} \mathbf{\Phi}_{T \setminus \hat{s}_1^i} \mathbf{x}_{T \setminus \hat{s}_1^i} + \phi_j' \mathbf{P}_{\hat{s}_1^i}^\bot \mathbf{v} \|_2 \label{eq:app:beta_lb12} \\
&\geq& \| \phi_j' \mathbf{\Phi}_{T \setminus \hat{s}_1^i} \mathbf{x}_{T \setminus \hat{s}_1^i} \|_2 - \| \phi_j' \mathbf{P}_{\hat{s}_1^i} \mathbf{\Phi}_{T \setminus \hat{s}_1^i} \mathbf{x}_{T \setminus \hat{s}_1^i} \|_2 - \| \phi_j' \mathbf{P}_{\hat{s}_1^i}^\bot \mathbf{v} \|_2. \label{eq:app:beta_lb1}
\end{eqnarray}
%
where \eqref{eq:app:beta_lb130} is because $\|\phi_j' \mathbf{r}_{\hat{s}_1^i}\|_2 = \sqrt{|\phi_j' \mathbf{r}_{\hat{s}_1^i}|^2} = |\phi_j' \mathbf{r}_{\hat{s}_1^i}|$ and \eqref{eq:app:beta_lb1} is from the triangle inequality.
Since \eqref{eq:app:beta_lb1} is satisfied for any $j \in T \setminus \hat{s}_1^i$, we have
%
\begin{eqnarray}
\label{eq:app:beta_lb2}
\| \phi_j' \mathbf{\Phi}_{T \setminus \hat{s}_1^i} \mathbf{x}_{T \setminus \hat{s}_1^i} \|_2
&\geq& \sqrt{1 - \delta_M} \sqrt{1 - \delta_{|T \setminus \hat{s}_1^i|}} \| \mathbf{x}_{T \setminus \hat{s}_1^i} \|_2, \label{eq:app:beta_lb211} \\
\| \phi_j' \mathbf{P}_{\hat{s}_1^i} \mathbf{\Phi}_{T \setminus \hat{s}_1^i} \mathbf{x}_{T \setminus \hat{s}_1^i} \|_2
&=& \| \phi_j' \mathbf{\Phi}_{\hat{s}_1^i} (\mathbf{\Phi}_{\hat{s}_1^i}'\mathbf{\Phi}_{\hat{s}_1^i})^{-1} \mathbf{\Phi}_{\hat{s}_1^i}' \mathbf{\Phi}_{T \setminus \hat{s}_1^i} \mathbf{x}_{T \setminus \hat{s}_1^i} \|_2 \label{eq:app:beta_lb2123} \\
&\leq& \delta_{|\hat{s}_1^i|+1} \| (\mathbf{\Phi}_{\hat{s}_1^i}'\mathbf{\Phi}_{\hat{s}_1^i})^{-1} \mathbf{\Phi}_{\hat{s}_1^i}' \mathbf{\Phi}_{T \setminus \hat{s}_1^i} \mathbf{x}_{T \setminus \hat{s}_1^i} \|_2 \label{eq:app:beta_lb2122} \\
&\leq& \frac{ \delta_{|\hat{s}_1^i|+1} }{ 1 - \delta_{|\hat{s}_1^i|} } \| \mathbf{\Phi}_{\hat{s}_1^i}' \mathbf{\Phi}_{T \setminus \hat{s}_1^i} \mathbf{x}_{T \setminus \hat{s}_1^i} \|_2 \label{eq:app:beta_lb2121} \\
&\leq& \frac{ \delta_{|\hat{s}_1^i|+1} \delta_K }{ 1 - \delta_{|\hat{s}_1^i|} } \| \mathbf{x}_{T \setminus \hat{s}_1^i} \|_2 \label{eq:app:beta_lb212}
\end{eqnarray}
%
and
%
\begin{eqnarray}
\| \phi_j' \mathbf{P}_{\hat{s}_1^i}^\bot \mathbf{v} \|_2
&\leq& \sqrt{1+\delta_M} \| \mathbf{P}_{\hat{s}_1^i}^\bot \mathbf{v} \|_2 \label{eq:app:beta_lb221} \\
&\leq& \sqrt{1+\delta_M} \| \mathbf{v} \|_2. \label{eq:app:beta_lb22}
\end{eqnarray}
%
where \eqref{eq:app:beta_lb211} and \eqref{eq:app:beta_lb221} are from Definition \ref{def_rip}, \eqref{eq:app:beta_lb2122} and \eqref{eq:app:beta_lb212} are from Lemma \ref{lem_rip_disjoint_set}, and \eqref{eq:app:beta_lb2121} is from Lemma \ref{lem_consqc_rip}.
Finally, since \eqref{eq:app:beta_lb1} is satisfied for any $j \in T \setminus \hat{s}_1^i$, we have
%
\begin{eqnarray}
\beta^i &=& \mathop{\min} \limits_{j \in T \setminus \hat{s}_1^i} |\phi_j' \mathbf{r}_{\hat{s}_1^i}| \\
&\geq& \| \phi_j' \mathbf{\Phi}_{T \setminus \hat{s}_1^i} \mathbf{x}_{T \setminus \hat{s}_1^i} \|_2 - \| \phi_j' \mathbf{P}_{\hat{s}_1^i} \mathbf{\Phi}_{T \setminus \hat{s}_1^i} \mathbf{x}_{T \setminus \hat{s}_1^i} \|_2 - \| \phi_j' \mathbf{P}_{\hat{s}_1^i}^\bot \mathbf{v} \|_2 \\
&\geq& \left( \sqrt{1 - \delta_M} \sqrt{1 - \delta_{|T \setminus \hat{s}_1^i|}} - \frac{ \delta_{|\hat{s}_1^i|+1} \delta_K }{ 1 - \delta_{|\hat{s}_1^i|} } \right) \| \mathbf{x}_{T \setminus \hat{s}_1^i} \|_2 - \sqrt{1+\delta_M} \| \mathbf{v} \|_2 \label{eq:app:beta_lb31}
 \\
&\geq& \left( 1 - \delta_M - \frac{ \delta_{K+1} \delta_K }{ 1 - \delta_K } \right) \| \mathbf{x}_{T \setminus \hat{s}_1^i} \|_2 - \sqrt{1+\delta_M} \| \mathbf{v} \|_2 \label{eq:app:beta_lb3}
\end{eqnarray}
%
where \eqref{eq:app:beta_lb3} is from Lemma \ref{lem_monotone_rip}.
%
\section{Proof of Lemma \ref{lem_alpha_ub}}
\label{app:alpha_ub}

Suppose $\hat{s}_1^i \subset T$ and let $\alpha^i = \mathop{\max} \limits_{j \in T^c} |\phi_j' \mathbf{r}_{\hat{s}_1^i}|$, then
%
\begin{eqnarray}
|\phi_j' \mathbf{r}_{\hat{s}_1^i}|
&=& \| \phi_j' \mathbf{r}_{\hat{s}_1^i} \|_2 \label{eq:app:alpha_ub130} \\
&=& \| \phi_j' \mathbf{P}_{\hat{s}_1^i}^\bot \mathbf{y} \|_2 = \| \phi_j' \mathbf{P}_{\hat{s}_1^i}^\bot (\mathbf{\Phi}_T \mathbf{x}_T + \mathbf{v}) \|_2 \label{eq:app:alpha_ub13}
\\
&=& \| \phi_j' \mathbf{P}_{\hat{s}_1^i}^\bot \mathbf{\Phi}_{T \setminus \hat{s}_1^i} \mathbf{x}_{T \setminus \hat{s}_1^i} + \phi_j' \mathbf{P}_{\hat{s}_1^i}^\bot \mathbf{v} \|_2 \label{eq:app:alpha_ub12}
\\
&=& \| \phi_j' \mathbf{\Phi}_{T \setminus \hat{s}_1^i} \mathbf{x}_{T \setminus \hat{s}_1^i} - \phi_j' \mathbf{P}_{\hat{s}_1^i} \mathbf{\Phi}_{T \setminus \hat{s}_1^i} \mathbf{x}_{T \setminus \hat{s}_1^i} + \phi_j' \mathbf{P}_{\hat{s}_1^i}^\bot \mathbf{v} \|_2 \label{eq:app:alpha_ub11}
\\
&\leq& \| \phi_j' \mathbf{\Phi}_{T \setminus \hat{s}_1^i} \mathbf{x}_{T \setminus \hat{s}_1^i} \|_2 + \| \phi_j' \mathbf{P}_{\hat{s}_1^i} \mathbf{\Phi}_{T \setminus \hat{s}_1^i} \mathbf{x}_{T \setminus \hat{s}_1^i} \|_2 + \| \phi_j' \mathbf{P}_{\hat{s}_1^i}^\bot \mathbf{v} \|_2 \label{eq:app:alpha_ub1}
\end{eqnarray}
%
where \eqref{eq:app:alpha_ub130} is because $\|\phi_j' \mathbf{r}_{\hat{s}_1^i}\|_2 = \sqrt{|\phi_j' \mathbf{r}_{\hat{s}_1^i}|^2} = |\phi_j' \mathbf{r}_{\hat{s}_1^i}|$, $\mathbf{P}_{\hat{s}_1^i} = \mathbf{\Phi}_{\hat{s}_1^i} (\mathbf{\Phi}_{\hat{s}_1^i}' \mathbf{\Phi}_{\hat{s}_1^i})^{-1} \mathbf{\Phi}_{\hat{s}_1^i}'$ in \eqref{eq:app:alpha_ub11}, $\mathbf{P}_{\hat{s}_1^i}^\bot = \mathbf{I} - \mathbf{P}_{\hat{s}_1^i}$, and \eqref{eq:app:alpha_ub1} follows the triangle inequality.
Since $j \in T^c$, we have
%
\begin{eqnarray}
\label{eq:app:alpha_ub21}
\| \phi_j' \mathbf{\Phi}_{T \setminus \hat{s}_1^i} \mathbf{x}_{T \setminus \hat{s}_1^i} \|_2
&\leq& \delta_{|T \setminus \hat{s}_1^i|+1} \| \mathbf{x}_{T \setminus \hat{s}_1^i} \|_2, \label{eq:app:alpha_ub211}\end{eqnarray}
%
\begin{eqnarray}
\| \phi_j' \mathbf{P}_{\hat{s}_1^i} \mathbf{\Phi}_{T \setminus \hat{s}_1^i} \mathbf{x}_{T \setminus \hat{s}_1^i} \|_2
&=& \| \phi_j' \mathbf{\Phi}_{\hat{s}_1^i} (\mathbf{\Phi}_{\hat{s}_1^i}'\mathbf{\Phi}_{\hat{s}_1^i})^{-1} \mathbf{\Phi}_{\hat{s}_1^i}' \mathbf{\Phi}_{T \setminus \hat{s}_1^i} \mathbf{x}_{T \setminus \hat{s}_1^i} \|_2 \label{eq:app:alpha_ub2123}\\
&\leq& \delta_{|\hat{s}_1^i| + 1} \| (\mathbf{\Phi}_{\hat{s}_1^i}'\mathbf{\Phi}_{\hat{s}_1^i})^{-1} \mathbf{\Phi}_{\hat{s}_1^i}' \mathbf{\Phi}_{T \setminus \hat{s}_1^i} \mathbf{x}_{T \setminus \hat{s}_1^i} \|_2 \label{eq:app:alpha_ub2122}\\
&\leq& \frac{ \delta_{|\hat{s}_1^i| + 1} }{ 1 - \delta_{|\hat{s}_1^i|} } \| \mathbf{\Phi}_{\hat{s}_1^i}' \mathbf{\Phi}_{T \setminus \hat{s}_1^i} \mathbf{x}_{T \setminus \hat{s}_1^i} \|_2 \label{eq:app:alpha_ub2121}\\
&\leq& \frac{ \delta_{|\hat{s}_1^i| + 1} \delta_K }{ 1 - \delta_{|\hat{s}_1^i|} } \| \mathbf{x}_{T \setminus \hat{s}_1^i} \|_2. \label{eq:app:alpha_ub212}
\end{eqnarray}
%
Also,
%
\begin{eqnarray}
\| \phi_j' \mathbf{P}_{\hat{s}_1^i}^\bot \mathbf{v} \|_2
&=& \sqrt{1+\delta_M} \| \mathbf{P}_{\hat{s}_1^i}^\bot \mathbf{v} \|_2 \label{eq:app:alpha_ub221} \\
&=& \sqrt{1+\delta_M} \| \mathbf{v} \|_2.\label{eq:app:alpha_ub22}
\end{eqnarray}
%
Using \eqref{eq:app:alpha_ub211}, \eqref{eq:app:alpha_ub212}, and \eqref{eq:app:alpha_ub22}, we have
%
\begin{eqnarray}
\label{eq:app:alpha_ub3}
|\phi_j' \mathbf{r}_{\hat{s}_1^i}|
&\leq& \| \phi_j' \mathbf{\Phi}_{T \setminus \hat{s}_1^i} \mathbf{x}_{T \setminus \hat{s}_1^i} \|_2 + \| \phi_j' \mathbf{P}_{\hat{s}_1^i} \mathbf{\Phi}_{T \setminus \hat{s}_1^i} \mathbf{x}_{T \setminus \hat{s}_1^i} \|_2 + \| \phi_j' \mathbf{P}_{\hat{s}_1^i}^\bot \mathbf{v} \|_2 \\
&\leq& \left( \delta_{|T \setminus \hat{s}_1^i|+1} + \frac{ \delta_{|\hat{s}_1^i| + 1} \delta_K }{ 1 - \delta_{|\hat{s}_1^i|} } \right) \| \mathbf{x}_{T \setminus \hat{s}_1^i} \|_2 + \sqrt{1+\delta_M} \| \mathbf{v} \|_2 \\
&\leq& \left( \delta_{K+1} + \frac{ \delta_{K + 1} \delta_K }{ 1 - \delta_K } \right) \| \mathbf{x}_{T \setminus \hat{s}_1^i} \|_2 + \sqrt{1+\delta_M} \| \mathbf{v} \|_2.
\end{eqnarray}
%

\bibliographystyle{IEEEbib}

\bibliography{IEEEabrv,CS_refs_a}

\end{document}